\documentclass[12pt]{article}
\usepackage{latexsym}
\usepackage{amsmath}
\usepackage{amssymb}
\usepackage{epsfig,graphics}
\usepackage{graphicx}
\usepackage{booktabs}
\usepackage{multirow}
\usepackage{caption}
\usepackage{subcaption}
\usepackage{tikz}
\usepackage{fixmath}
\usetikzlibrary{matrix,snakes,arrows,shapes,decorations.pathmorphing,decorations.markings,calc}

\newcommand{\be}{\begin{equation}}
\newcommand{\ee}{\end{equation}}

\begin{document}

\begin{titlepage}

\vspace*{0.6in}
 
\begin{center}
{\large\bf Pfaffian particles and strings in SO(2N) gauge theories}\\
\vspace*{0.75in}
{Michael Teper \\
\vspace*{.25in}
Rudolf Peierls Centre for Theoretical Physics,\\ Clarendon Laboratory,University of Oxford,\\
Parks Road, Oxford OX1 3PU, UK\\
\centerline{and}
All Souls College, University of Oxford,\\
High Street, Oxford OX1 4AL, UK}
\end{center}

\vspace*{0.4in}

\begin{center}
{\bf Abstract}
\end{center}

We introduce (generalised) Pfaffian operators into our lattice calculations of
the mass spectra and confining string tensions of $SO(2N)$ gauge theories,
complementing the conventional trace operators used in previous lattice calculations.
In $SO(6)$ the corresponding `Pfaffian' particles match the negative
charge conjugation particles of $SU(4)$, thus resolving a puzzle arising
from the observation that $SO(6)$ and $SU(4)$ have the same Lie algebra.
The same holds true (but much more trivially) for $SO(2)$ and $U(1)$. For
$SO(4)$ the Pfaffian particles are degenerate with, but orthogonal to, those
obtained with the usual single trace operators. That is to say, there is a
doubling of the spectrum, as one might expect given that the Lie algebra of
$SO(4)$ is the same as that of $SU(2)\times SU(2)$. Additional  $SO(8)$
and  $SO(10)$ calculations of the Pfaffian spectrum confirm the naive expectation
that these masses increase with $N$, so that they cease to play a role in the
physics of $SO(N)$ gauge theories as $N\to\infty$. We also calculate the
energies of Pfaffian `strings' in these gauge theories. Although all our lattice
calculations are for gauge theories in $D=2+1$, similar conclusions
should hold for $D=3+1$.

\vspace*{0.95in}

\leftline{{\it E-mail:} mike.teper@physics.ox.ac.uk}

\end{titlepage}

\setcounter{page}{1}
\newpage
\pagestyle{plain}

\tableofcontents

\section{Introduction}
\label{section_intro}

We begin with our original motivation for the calculations of this paper.
We recall that certain pairs of $SU(N)$ and $SO(N^\prime)$ gauge theories share
the same Lie algebra. These pairs are $SO(3)$ and $SU(2)$, $SO(4)$ and
$SU(2)\times SU(2)$, and $SO(6)$ and $SU(4)$. Whether the differing global properties
of the groups in each pair affect the physics is an interesting question
that has provided one of the motivations for recent calculations
\cite{RLMT_SON_a,RLMT_SON_b,MT_SON}
of the low-lying mass spectra of $SO(N)$ gauge theories.
For technical reasons (to do with the location of the transition
between strong and weak coupling physics) these have been performed in $2+1$
rather than $3+1$ dimensions. These spectra have then been compared to existing
$SU(N)$ calculations in $D=2+1$
\cite{AAMT_SUN}.
Since $SO(N)$ is real, one might naively assume that there is no room for
negative charge conjugation states so that, for example, there are no
states in the $SO(6)$ spectrum that correspond to the $C=-$ states of $SU(4)$.
And indeed one finds that if one uses the standard single (or multi) trace
operators this is so
\cite{RLMT_SON_a,RLMT_SON_b,MT_SON}.
Moreover the lightest $C=+$ spectra do appear to be
consistent between the corresponding pairs of $SU(N)$ and $SO(N^\prime)$
gauge theories
\cite{RLMT_SON_a,RLMT_SON_b,MT_SON}.
This is also the case for the planar large-$N$ limits of $SO(N)$ and
$SU(N)$ gauge theories, as predicted by the usual diagrammatic large-$N$
counting
\cite{largeN_SUN,largeN_SON}.
However the fact that the light $C=-$ states of $SU(4)$ do not appear
to be encoded in the corresponding $SO(6)$ spectrum creates a puzzle, as
emphasised in
\cite{MT_spinorial}.
For example, a sufficiently excited $C=+$ glueball in $SU(4)$ can decay
into two $C=-$ glueballs. This will contribute to its decay width (and
will shift its mass). If the corresponding $C=+$ glueball in the  $SO(6)$
theory is identical, then its decay products will include states composed
of these two $C=-$ glueballs arbitrarily far apart. This is hard to
understand if the theory does not include single particles corresponding
to the $C=-$ glueballs of $SU(4)$. One can observe a similar puzzle
concerning flux tubes in the fundamental representation of $SU(4)$, and
hence in the spinorial of $SO(6)$, as described in detail in
\cite{MT_spinorial}.
In this paper we shall focus on resolving the glueball puzzle, leaving
the flux tube puzzle to future work.

As we shall show below, the solution to the above glueball puzzle is to
be found in the fact that the $SO(6)$ gauge theory possesses an additional type 
of gauge invariant operator that is orthogonal to the usual trace operators.
This is a generalisation of the Pfaffian operator for $SO(2N)$ gauge theories
whose encoding in the AdS/CFT correspondence has been discussed in
\cite{EW_98}.
Such operators were not included in the calculations of  
\cite{RLMT_SON_a,RLMT_SON_b}
and so the $SO(2N)$ spectra obtained therein are incomplete, albeit correct as
far as they go. This operator plays no role in $SU(N)$ gauge theories,
for reasons discussed below. In this paper we shall calculate the masses
of the lightest `Pfaffian particles' (following the nomenclature of
\cite{EW_98})
in a number of $SO(2N)$ gauge theories and we will show that this resolves
a number of puzzles including that of the missing `$C=-$' states in $SO(6)$.
We will include in our calculations the pair $SO(2)$ and $U(1)$, where one
can also ask how the $C=-$ states of $U(1)$ are encoded in $SO(2)$. This case
has the advantage of being so trivial that one can immediately see how
the `Pfaffian particles' resolve the puzzle. While most of our calculations
will be for $SO(2)$, $SO(4)$ and $SO(6)$, for the reasons outlined above we
will also perform calculations for $SO(8)$ and $SO(10)$ so as to see what
happens to the `Pfaffian particles' as $N\to\infty$.

In our calculations there are a number of properties of the `Pfaffian' operators
that are important for us. In general we shall provide numerical rather than
analytic evidence for these properties, except where we are aware of simple analytic
arguments. These properties are discussed in Section~\ref{section_Pfaffian}.
Prior to that, in Section~\ref{section_lattice}, we  outline
how the lattice calculations are performed. Here we address 
the caveats concerning the accuracy of our mass calculations; it is important
to bear these in mind later on in the paper when we compare the spectra
of the corresponding $SO(N)$ and $SU(N^\prime)$ gauge theories.
In Section~\ref{section_spectrum} we perform calculations in $SO(2)$, $SO(4)$
and $SO(6)$, and compare with the results of calculations in
$U(1)$,  $SU(2) \times SU(2)$ and  $SU(4)$ respectively. The latter calculations
are carried out at bare couplings such that the mass gap is nearly the same
within each  $SO(N)$ and $SU(N^\prime)$ pair.
This allows for the comparisons to be reasonably direct. 
We also perform calculations in $SO(8)$ and $SO(10)$ and use our various
results to make plausible extrapolations in $N$. In this exploratory
study we do not attempt to extrapolate our results to the continuum
limit of the lattice gauge theories but instead choose bare
couplings where earlier work, in both $SO(N)$ and $SU(N)$, has shown
lattice corrections to be very small for the masses calculated. And
we assume that the same is true for the Pfaffian particles.
In addition to particle masses
one can also take the Pfaffian of a non-contractible loop that winds around
a spatial circle. The trace of such a loop projects onto a flux loop that
winds around the circle and from its energy we can estimate the confining
string tension. In Section~\ref{section_confinement} we ask whether this is
also true of the  Pfaffian of such a loop and if so what is the flux that 
it carries. Since the present work is exploratory in nature (albeit with a
number of interesting conclusions) there are various open questions that
we have encountered but have not tried to  address because of our
focus on resolving the puzzles listed above. We point to
a number of these questions in Section~\ref{section_questions} and
discuss different ways they may be resolved. Finally in our concluding 
Section~\ref{section_conclusions} we summarise our results.

\section{Lattice preliminaries}
\label{section_lattice}

Our lattice calculations are standard and we refer to
\cite{AAMT_SUN}
for a detailed description of the $SU(N)$ calculations, and to
\cite{RLMT_SON_a,RLMT_SON_b}
for details of the $SO(N)$ calculations.

\subsection{path integral}
\label{subsection_pathintegral}

Our Euclidean space-time is a finite cubic lattice, with a lattice spacing
denoted by $a$ and a size $l_x\times l_y\times l_t$ in lattice units.
The boundary conditions for the fields are periodic.
For $SO(N)$ our degrees of freedom are $N\times N$ real orthogonal matrices
with unit determinant, which are assigned to the links $l$ of the lattice.
For $SU(N)$ the matrices are $N\times N$ complex
unitary matrices with unit determinant. The matrix on the link $l$ will
normally be denoted by $U_l$, if forward going, and $U_l^\dagger$ if
backward going, although we will sometimes denote $SO(N)$ matrices by $O_l$. We will
sometimes write $U_l$ as $U_\mu(n)$ where $\mu$ is the direction of the link
and $n$ is an integer (or triplet of integers) labelling the site from which
the link emanates in a forward going direction. The partition function is 
\begin{equation}
Z=\int {\cal{D}}U \exp\{-\beta S[U]\}
\label{eqn_Z}
\end{equation}
where ${\cal{D}}U$ is the group Haar measure and we use the standard plaquette action
for $S[U]$, 
\begin{equation}
  S[U]= \frac{1}{2}\sum_{\mu\neq\nu,n}\left\{1 - \frac{1}{N}\mathrm{Tr} \{U_\mu(n)U_\nu(n+\hat{\mu})
  U_\mu^\dagger(n+\hat{\nu}) U_\nu^\dagger(n)\} \right\}
\label{eqn_S}
\end{equation}
where for $SO(N)$ we obviously have  $O_l^\dagger=O_l^T$. When we take
the (naive) continuum limit and compare to the continuum action we find
that $\beta=2N/ag^2$. We recall that in $D=2+1$ $g^2$ has dimensions of mass,
so that $lg^2$ is the dimensionless running coupling on the length scale $l$.
On the lattice the degrees of freedom are defined on the length scale of the
lattice spacing $a$ and so $ag^2$ is the appropriate dimensionless running coupling.

We will use the standard plaquette action in all our calculations in this
paper.

\subsection{energies from correlators}
\label{subsection_energies}

We calculate the mass spectrum from correlators of gauge invariant operators.
Suppose $\psi(t)$ is a gauge invariant operator with some specific $J^{PC}$
quantum numbers and with zero momentum, $p=0$. Then
\begin{equation}
<\psi^\dagger(t) \psi(0)> 
= \sum_n |<n|\psi|vac>|^2 e^{-E_n t} 
\stackrel{t\to\infty}{=}|<0|\psi|vac>|^2 e^{-Mt} 
\label{eqn_M}
\end{equation}
where $M$ is the mass of the lightest state with the quantum numbers of the
operator $\psi$, and $|0\rangle$ is the corresponding state.
The operator $\psi$ will typically be a linear combination
of more elementary operators. This will typically be based on a path
ordered product
of link matrices around some closed path $\mathcal{C}$ that starts and ends at
some site $x$ (shorthand for $\{x,y,t\}$). Call it $\Phi_{\mathcal{C}}$. In $SO(N)$
and $SU(N)$ this will be an $N\times N$ matrix. To make it gauge invariant
one can take the trace, $\mathrm{Tr}\{\Phi_{\mathcal{C}}(x)\}$. (One can also use
products of traces, which will have a larger overlap onto multiparticle states.)
For $SU(N)$ traced operators provide a systematic way to calculate the full
spectrum of the theory. This same technique was taken over in
\cite{RLMT_SON_a,RLMT_SON_b}
to calculate the $SO(N)$ mass spectrum but in that case it turns out to be
incomplete, in an interesting way, as we shall see below.

In practice we use a large basis of such closed loop operators, which include
operators that are smooth on physical length scales, and hence large in lattice
units, so as to have a good overlap onto the lightest physical states.
(This can be efficiently achieved by iteratively `blocking' the link matrices
\cite{block_a,block_b}.)
A good overlap is crucial since the numerical calculations of a correlator 
$<\psi^\dagger(t) \psi(0)>$ will have statistical errors roughly independent
of $t$ while the interesting physical `signal' in eqn(\ref{eqn_M}) decreases
exponentially in $t$. So the lightest state needs to emerge from the background
of excited states at small $t$ if it is not to be drowned in the statistical
noise. This requires that the normalised overlap
$<0|\psi|vac>|^2/<\psi^\dagger(0) \psi(0)>$ should be not not very small,
and in practice one finds that it needs to be larger than $\sim 0.5$
if we are to capture useful information about the state. To obtain the
linear combination of loop operators that `best' approximates the wavefunction
of the lowest eigenstate we apply a variational calculation based on
maximising the transfer matrix, i.e. $\exp\{-aH\}$ in continuum
language. This gives us an approximate ground state wavefunctional
$\Psi_0$ and we then calculate the correlator
$<\Psi_0^\dagger(t) \Psi_0(0)>$ from which we extract our mass estimate.
Here it is convenient to define an effective mass by
\begin{equation}
  \exp\{-aM_{eff}(t)\}
  =
  \frac{<\Psi^\dagger(t+a) \Psi(0)>}{ <\Psi^\dagger(t) \Psi(0)>}.
\label{eqn_Meff}
\end{equation}
If for some $t\geq t_o$  the correlator is dominated by a single exponential,
then $aM_{eff}(t)$ will become independent of $t$ for $t\geq t_o$ and will be
equal to the desired lightest mass. Thus we look for a `plateau' (within errors)
in the effective mass, from which we extract an estimate of the true mass.
Since the errors are roughly independent of $t$, the error on $aM_{eff}(t)$
will grow exponentially with $t$ and if $aM$ is large then there will be
some guesswork involved in identifying the effective mass plateau. The typical
error is to extract $aM$ from  $aM_{eff}(t)$ at too small a value of $t$,
where it still receives significant contributions from heavier excited states.
This will obviously lead to an overestimate of the mass. Moreover, since
the error on  $aM_{eff}(t)$ increases with $t$, our statistical error
on this overestimated mass will be smaller than it should be.  All this
can also occur if the mass is not large but the overlap is small. Since the
credibility of our conclusions will ultimately rest on the reliability
of our masses, we will provide the explicit examples of relevant
effective mass plots at appropriate points in the paper.

Once we have $\Psi_0$ we can repeat the variational procedure in a
basis orthogonal to $\Psi_0$ and this will give us an approximate
wavefunctional$\Psi_1$ for the first excited state, from whose
correlator we can extract an approximation to the energy of the
excited state. We can repeat the process for higher exctied states.

Normally the state that is picked out by the variational procedure as
the candidate ground state does indeed have the largest overlap
onto the true ground state. However if the lightest state has a poor
overlap onto our basis of operators, then it may appear in the
larger-$t$ tail of the correlator of an excited state. We shall see
an example of this later on in the paper.

To obtain glueball masses with particular quantum numbers such as spin $J$
and parity $P$ we use operators with those same symmetries. Note that
the limited $\pi/2$ rotational symmetry means that the `$J=0$' representation
of the 2D rotation group contains states that become $J=4,8,...$ in the
continuum limit. Similarly  for `$J=2$' and `$J=1$'. For convenience we
ahall, from now on, label all the states by $J=0$ or $2$ or $1$.
The parity reflections in $x$ and $y$ can be rotated to each other,
but there is a further parity in the $x=y$ axis that cannot be --
although it can be, of course, in the continuum limit. Our parity
will always be in the $x$ or $y$ axis.
Note also that for $J\neq 0$ we have parity doubling: this is exact 
for $J=1^\pm$ but for $J=2^\pm$ can be broken by finite volume or
sub-leading lattice corrections. For a more detailed discussion see
\cite{HMMT_a,HMMT_b}.

If the theory is confining then there is a finite volume state
consisting of a flux tube wrapped around a spatial circle. If the
length of the circle is $l$ and the ground state energy is $E(l)$
then we extract the string tension $\sigma$ using
\begin{equation}
  E(l) = \sigma l \left(1-\frac{\pi}{3\sigma l^2}\right)^{\frac{1}{2}} 
\label{eqn_NG}
\end{equation}
This expression is a very good approximation down to very small $l$
\cite{string_LAT_a,string_LAT_b,string_TH_a,string_TH_b,string_TH_c}.
We extract $E(l)$ using operators that consist of products of link
matrices around a curve that winds once around the spatial circle.
This is a non-contractible loop in contrast to the contractible loops
appropriate for glueball operators. For traced operators one can
show that such a non-contractible loop operator has no overlap
on glueballs for those theories with a non-trivial centre symmetry,
such as $SU(N)$ and $SO(2k)$, if we are in the confining phase.

\section{Some properties of `Pfaffian' operators in SO(2N)}
\label{section_Pfaffian}

We begin with the standard definition of the Pfaffian and how it usually 
appears in gauge theories. We then move onto generalisations for both
glueballs and winding flux tubes. To be useful in calculations of energies
these operators must be gauge invariant as we shall see they
are. In addition, to provide a useful addition to the standard trace operators
used in previous lattice calculations
\cite{RLMT_SON_a,RLMT_SON_b,MT_SON},
they should project onto states with which the trace operators have no, or very
little, overlap. We shall show, numerically, that this is indeed the case for
the $SO(N)$ groups that we consider in this paper.

\subsection{the Pfaffian operator}
\label{subsection_Pfaffian}

Let $A$ be a $2k\times 2k$ skew-symmetric matrix, i.e. $A^T=-A$. Then
its Pfaffian, $\mathrm{Pf}(A)$, is defined to be
\begin{equation}
  \mathrm{Pf}(A) = \frac{1}{2^kk!} \epsilon_{i_1i_2...i_{2k}}
  A_{i_1i_2}A_{i_3i_4}  ... A_{i_{2k-1}i_{2k}}
  \label{eqn_PfA}
\end{equation}
where $ \epsilon_{i_1i_2...i_{2k}}$ is the totally antisymmetric (Levi-Civita)
tensor with $2k$ indices. (We do not distinguish upper and lower indices.)
Moreover for such a matrix 
\begin{equation}
  \mathrm{Pf}(A)^2 = \mathrm{det}(A)
  \label{eqn_detA}
\end{equation}
Recall that the generators of $SO(N)$ are skew-symmetric. So if we choose
$A$ to be  an antisymmetric second rank tensor transforming in 
the adjoint of $SO(2k)$, e.g. the field strength, then $\mathrm{Pf}(A)$
will be invariant under a gauge transformation $V(x)$ since it
will satisfy eqn(\ref{eqn_detA}) and $\mathrm{det}(A)$ is gauge invariant:
$\mathrm{det}(A(x))\stackrel{V(x)}{\to}\mathrm{det}(V(x)A(x)V^T(x)) =
\mathrm{det}(A(x))$ since $\mathrm{det}(V) = 1$. To be more precise,
in order to evade the $\pm$ ambiguity in taking the square root of
$\mathrm{det}(A)$ in eqn(\ref{eqn_detA}) we should restrict ourselves
to gauge transformations $V(x)$ that are continuously connected to the
identity. Along such a path the Pfaffian will not change sign by continuity.
These are the usual `small' gauge tranformations in contrast
to the `large' gauge transformations that are typically associated
with topological fluctuations in one higher Euclidean dimension.
(In fact a particular `large' transformation under which the Pfaffian
flips sign will play a role in our arguments later on in this paper.)

Following
\cite{EW_98}
we will refer to the states that this Pfaffian (and its generalisations
below) couple to as `Pfaffian particles'.
Naively these particles will be composed of $k$ gauge bosons in $SO(2k)$ and so one
might expect that they will become heavy as $k$ grows and will become irrelevant
to the light particle spectrum of $SO(2k)$ as $k\to\infty$. Although this is an
expectation that we will test later on in this paper, our primary interest is
in small $k$ where these states should be important.

\subsection{generalised Pfaffians}
\label{subsection_generalisedPfaffian}

On the lattice we do not work with the gluon fields directly but rather with
$SO(N)$ matrices $\Phi(\mathcal{C};x)$ that are obtained by multiplying the link
matrices around some path $\mathcal{C}$ that begins and ends at the
site $x$. (In terms of continuum fields, this becomes just the path ordered
exponential of the gauge fields around  the contour $\mathcal{C}$.)
We define the Pfaffian $\mathrm{Pf}(\Phi)$ of such a matrix $\Phi$ by replacing
$A$ with $\Phi$ in eqn(\ref{eqn_PfA}):
\begin{equation}
  \mathrm{Pf}(\Phi(x)) = \frac{1}{2^k k!} \epsilon_{i_1i_2...i_{2k}}
  \Phi(x)_{i_1i_2}\Phi(x)_{i_3i_4}  ... \Phi(x)_{i_{2k-1}i_{2k}}
  \label{eqn_PfPhi}
\end{equation}
This is a generalisation of the usual definition of a Pfaffian in eqn(\ref{eqn_PfA})
since the $SO(N)$ matrix  $\Phi(\mathcal{C};x)$ is not skew symmetric, but rather 
satisfies $\Phi^T=\Phi^{-1}$. Such an operator will only be useful if it is gauge
invariant which we need to show because the relation in eqn(\ref{eqn_detA}) that
we used to show gauge invariance will no longer hold in general. So consider
an $SO(N)$  gauge transformation $V(x)$ under which $\Phi(x)\to V(x)\Phi(x)V^T(x)$.
Then the Pfaffian of $\Phi$ changes as
\begin{eqnarray}
  2^k k! \mathrm{Pf}(\Phi(x))
  & = & \epsilon_{i_1i_2...i_{2k}} \Phi_{i_1i_2} ... \Phi_{i_{2k-1}i_{2k}}  \nonumber \\
  & \stackrel{V}{\to} & 2^k k! \mathrm{Pf}(V\Phi(x)V^T)  \nonumber \\
  & = & \epsilon_{i_1i_2...i_{2k}} V_{i_1,j_1}\Phi_{j_1j_2} V^T_{j_2,i_2} ...
  V_{i_{2k-1}j_{2k-1}}\Phi_{j_{2k-1}j_{2k}}V^T_{j_{2k}i_{2k}}  \nonumber \\
  & = & \epsilon_{i_1i_2...i_{2k}} V_{i_1,j_1}\Phi_{j_1j_2} V_{i_2,j_2} ...
  V_{i_{2k-1}j_{2k-1}}\Phi_{j_{2k-1}j_{2k}}V_{i_{2k}j_{2k}}  \nonumber \\
  & = & \epsilon_{i_1i_2...i_{2k}} V_{i_1,j_1}V_{i_2,j_2} ... V_{i_{2k-1}j_{2k-1}}V_{i_{2k}j_{2k}} 
  \Phi_{j_1j_2} ... \Phi_{j_{2k-1}j_{2k}}  \nonumber \\
  & = & \epsilon_{j_1j_2...j_{2k}} \Phi_{j_1j_2} ... \Phi_{j_{2k-1}j_{2k}}  \nonumber \\
  & = &  2^k k!  \mathrm{Pf}(\Phi(x))
  \label{eqn_PfVPhiV}
\end{eqnarray}
where in the fourth line we have used the fact that $V^T_{ab}=V_{ba}$, and in the
sixth line the identity
\begin{equation}
  \epsilon_{i_1i_2...i_N} V_{i_1,j_1}V_{i_2,j_2} ... V_{i_{N-1}j_{N-1}}V_{i_{N}j_{N}}
  =
  \epsilon_{j_1j_2...j_{N}}
  \label{eqn_PfV}
\end{equation}
for $SO(N)$ matrices $V$ which is valid for odd $N$ as well as for even $N$.
Thus our generalised Pfaffian is also gauge invariant and can be a useful operator
in lattice mass calculations.

From the above derivation it is clear that if we have matrices $\Phi(\mathcal{C}_i;x)$
that differ because they involve products of link matrices around different paths
$\mathcal{C}_i$, but with all the paths still begining and ending at the same point
$x$, then the even more general Pfaffian, defined by
\begin{equation}
  \mathrm{Pf}(\{\Phi(\mathcal{C}_i;x)\}) = \frac{1}{2^k k!} \epsilon_{i_1i_2...i_{2k}}
  \Phi(\mathcal{C}_1;x)_{i_1i_2}\Phi(\mathcal{C}_2;x)_{i_3i_4}  ...
  \Phi(\mathcal{C}_k;x)_{i_{2k-1}i_{2k}},
  \label{eqn_PfPhiCl}
\end{equation}
is also gauge invariant. So this provides an even more extensive set of operators that
one can use in mass calculations.

Another more illuminating way to see the gauge invariance of an adjoint operator
$\Phi(x)$ is as follows. Let us decompose $\Phi$ into a sum of symmetric and
skew-symmetric pieces:
$\Phi = \Phi^e + \Phi^o$ where $\Phi^o_{ij}=-\Phi^o_{ji}$ and $\Phi^e_{ij}=\Phi^e_{ji}$.
Clearly $\Phi^o = 1/2(\Phi - \Phi^T)$ and  $\Phi^e = 1/2(\Phi + \Phi^T)$. (Aside: since
$\Phi$ is an $SO(N)$ group element, we can write  $\Phi = \exp\{A\}$ where $A$ is a
skew-symmetric Lie algebra element, and expanding the exponential we see that $\Phi^e$
and $\Phi^o$ are just the sums of even and odd powers of $A$ respectively.)
Inserting this sum for each occurrence of $\Phi$ into $\mathrm{Pf}(\Phi)$ we see
that any term containing at least one $\Phi^e$ must vanish. That is to say,
\begin{equation}
  \mathrm{Pf}(\Phi(x)) = \mathrm{Pf}(\Phi^e(x)+\Phi^o(x))= \mathrm{Pf}(\Phi^o(x)) =
  \{ \mathrm{det}(\Phi^o(x))\}^{\frac{1}{2}}
  \label{eqn_PfPhio}
\end{equation}
where we can now use eqn(\ref{eqn_detA}) since $\Phi^o$ is skew-symmetric.
Now under a gauge transformation $V(x)$, we have 
\begin{eqnarray}
  \mathrm{Pf}(\Phi(x)) \to  \mathrm{Pf}(V(x)\Phi(x)V^T(x)) & = & \mathrm{Pf}(V(x)\Phi^o(x)V^T(x)) \nonumber \\
  & = & \mathrm{det}(V(x)\Phi^o(x)V^T(x))^{\frac{1}{2}}  \nonumber \\
  & = & \mathrm{det}(\Phi^o(x))^{\frac{1}{2}}
  \label{eqn_PfPhiClb}
\end{eqnarray}
since $\mathrm{det}(V(x))=\mathrm{det}(V^T(x))=1$ and one can easily show
that the skew symmetric piece of
$V(x)\Phi(x)V^T(x)$ is just $V(x)\Phi^o(x)V^T(x)$. So we see that
$\mathrm{Pf}(\Phi(x))$ is indeed gauge invariant.
(Again, we are only considering `small' gauge transformations in order
to avoid a possible sign ambiguity in the square root.)

\subsection{Pfaffians and SU(N)}
\label{subsection_suNPfaffian}

As an aside, we remark that eqn(\ref{eqn_PfV}) also holds if $V$ is an $SU(N)$
matrix. However in the $SU(N)$ gauge theory a gauge transformation leads to 
$\Phi(x)_{ab}\to V(x)_{ac}\Phi(x)_{cd}V^{\dagger}(x)_{db}=V(x)_{ac}\Phi(x)_{cd}V^{\star T}(x)_{db}=V(x)_{ac}V^{\star}(x)_{bd}\Phi(x)_{cd}$
and so the Pfaffian of $\Phi$ is not gauge invariant since eqn(\ref{eqn_PfV})
does not hold if we replace every second $V$ in the product by its complex conjugate.
On the other hand, if we introduce fundamental fields $\psi(x)$ into the $SU(N)$
gauge theory, then eqn(\ref{eqn_PfV}) ensures that the operator
$\epsilon_{i_1i_2...i_{N}} \psi_{i_1} \psi_{i_2} ... \psi_{i_{N}}$ is gauge invariant.
If $\psi$ is a fermion, then this is of course just the gauge-invariant operator
for a baryon in $N$-colour QCD. The same holds if we introduce fundamental fields
into an $SO(N)$ gauge theory, for any value of $N$.

\subsection{Pfaffians for flux tubes}
\label{subsection_stringPfaffian}

The above demonstration of the gauge invariance of  $\mathrm{Pf}(\{\Phi(\mathcal{C};x)\})$
is clearly valid irrespective of whether the curve $\mathcal{C}$ is contractible or
not. Now, as described in Section~\ref{section_lattice}, the traces of contractible
loops form a basis for glueball operators while the traces of non-contractible
loops that wind once around a periodic spatial direction typically form a basis
for operators that project onto confining flux tubes that wrap once around that
spatial direction. This distinction can be readily motivated using the $Z_2$ centre
symmetry of the $SO(2k)$ gauge theory. If we multiply by $-1 \in Z_2$ the matrices
$U_{\mu=x}(n_x,n_y,n_t); \forall n_y,n_t$ then a contractible loop, label it
$\Phi_g$, will transform as $\Phi_g \to \Phi_g$, since the number of factors of
$-1$ will be even, but a non-contractible loop that winds once around the $x$-torus,
label it $\Phi_l$, will clearly transform as $\Phi_l \to -\Phi_l$ since the number
of factors of $-1$ will be odd. Under this field transformation the Haar measure
is invariant and so is the action since the plaquette (a contractible loop)  is
unchanged.
So the fields have the same weight and if the $Z_2$ symmetry is not spontaneously
broken we will necessarily have 
\begin{equation}
  \langle \mathrm{Tr}(\Phi_g(n))\mathrm{Tr}(\Phi_l(n^\prime)) \rangle =
  - \langle \mathrm{Tr}(\Phi_g(n))\mathrm{Tr}(\Phi_l(n^\prime)) \rangle
  = 0  \quad ; \quad \forall n,n^\prime,g,l.
  \label{eqn_corlg}
\end{equation}
That is to say, the states produced by contractible and non-contractible loops
are orthogonal to each other, as one would expect for glueballs and winding
flux tubes in a confining theory.

We can now apply the same argument to the Pfaffian of $\Phi_g$ and $\Phi_l$
in the $SO(2k)$ gauge theory.
From eqn(\ref{eqn_PfPhi}) we see that under this $Z_2$ field tranformation we
have $\mathrm{Pf}\{\Phi_g\} \to \mathrm{Pf}\{\Phi_g\}$ while
$\mathrm{Pf}\{\Phi_l\} \to (-1)^k \mathrm{Pf}\{\Phi_l\}$. Thus
$\mathrm{Pf}\{\Phi_x\}|vac\rangle$ is orthogonal to $\mathrm{Pf}\{\Phi_g\}|vac\rangle$
and also to  $\mathrm{Tr}(\Phi_g(x))|vac\rangle$ if $k$ is odd.
That is to say, we expect $\mathrm{Pf}\{\Phi_l\}$ to project onto confining
flux tubes that wind around the $x$-torus in $SO(4k+2)$ theories but not
necessarily in $SO(4k)$ theories. The former includes $SO(2),\, SO(6),\, SO(10)$
while the latter includes  $SO(4),\, SO(8)$.

For $SO(4k)$ theories we can instead use the more general Pfaffian in
eqn(\ref{eqn_PfPhiCl}). For instance for $SO(4)$ the operator
$\epsilon_{ijkl}\{\Phi_g(x)\}_{ij}\{\Phi_l(x)\}_{kl}$
is gauge invariant and changes sign under the centre transformation discussed
above, and so produces states orthogonal to glueball states. In general
this will be the case whenever the generalised Pfaffian contains an odd
number of winding loops with the remaining loops being contractible.

\subsection{Pfaffians: orthogonality}
\label{subsection_orthoPfaffian}

Clearly the above Pfaffian operators will only be really useful if they have large
projections onto states that our standard traced operators do not. In that case our
usual lattice calculation with traced operators will completely miss these states,
and the Pfaffian operators will be essential for their identification.

To see if this is the case or not we have performed numerical calculations
of overlaps of the form
\begin{equation}
  O_{IJ} = 
  \frac{\langle \mathrm{Tr}(\Phi_I(0))\mathrm{Pf}(\Phi_J(0)) \rangle}
  {\langle \mathrm{Tr}(\Phi_I(0))\mathrm{Tr}(\Phi_I(0)) \rangle^\frac{1}{2}
    \langle \mathrm{Pf}(\Phi_J(0))\mathrm{Pf}(\Phi_J(0)) \rangle^\frac{1}{2}}
    \label{eqn_TrPfover}
\end{equation}
for various loops labelled $I$ and $J$ and we have done so in all the
$SO(2k)$ theories analysed in this paper . In this exploratory analysis we have
not used all the loops in our glueball/flux tube calculations but only a
limited subset which nonetheless includes operators with a good overlap
onto the ground states (both Pfaffian and traced respectively). What we find
is that all the overlaps so tested are consistent with zero within very
small statistical errors. The implication (albeit based on a limited
numerical calculation)
is that the Pfaffian operators do project onto states that will be
numerically invisible to the traced operators, and that their use is therefore
essential for the identification of these states.

There is a simple argument to strengthen this conclusion. Consider an
$SO(2k)$ matrix $\Phi(\mathcal{C},n)$ obtained from the product of link
matrices around the contour $\mathcal{C}$ that is open at the site $n$.
As we have seen above we have $\mathrm{Pf}(\Phi)=\mathrm{Pf}(\Phi^o)$
where $\Phi^o = 1/2(\Phi - \Phi^T)$. Now we expect that the matrices $\Phi$
and $\Phi^T$ should be equally likely in the integration over all fields.
That is to say for each field that $\Phi$ and $\Phi^T$ take some values,
there is another field, with equal weight, for which they take values $\Phi^\prime$
and $\Phi^{\prime T}$ such that $\Phi^\prime = \Phi^T$ and $\Phi^{\prime T}=\Phi$.
Now clearly $\mathrm{Pf}(\Phi^{\prime}) = (-1)^k\mathrm{Pf}(\Phi)$, since the
Pfaffian contains a product of $k$ elements of $\Phi^o$. Since on the other hand
$\mathrm{Tr}(\Phi^{\prime})=\mathrm{Tr}(\Phi^T)=\mathrm{Tr}(\Phi)$ it
immediately follows that 
\begin{equation}
  \langle \mathrm{Tr}(\Phi(x))\mathrm{Pf}(\Phi(x)) \rangle = 0 \qquad,\quad k=\mathrm{odd}.
  \label{eqn_PfPhiClc}
\end{equation}
That is to say the operator $\mathrm{Pf}(\Phi)$ is exactly orthogonal to the
operator  $\mathrm{Tr}(\Phi)$ in $SO(2k)$ gauge theories when $k$ is odd. This adds
support to the idea that the spectrum divides into two sectors, one accessed by
using the traces of closed loops and the other by using the Pfaffians of these loops.

The above argument is, however, not enough to show that the spectrum divides into
two sectors. What we wish to do is to show that $\mathrm{Tr}(\tilde{\Phi})$ and
$\mathrm{Pf}(\Phi)$ are orthogonal for any two operators $\tilde{\Phi}$ and $\Phi$.
To show this the relevant observation is that, as discussed in Section 4.1 of
\cite{EW_98},
our $SO(N)$ lattice gauge theory is in fact symmetric under $O(N)$ and not
just $SO(N)$ and it is the $Z_2$ of $O(N)/SO(N)$ that provides the quantum number
dividing the spectrum into two sectors\footnote{I am grateful to the referee
for pointing this out to me.}. It is worth being more specific here.
$O(N)$ differs from $SO(N)$ by the inclusion of elements $V$ with
$\mathrm{det}(V)=-1$. To generate these  elements consider the matrix $O_p$
which is the unit matrix except that the $(1,1)$ element is $-1$. This
matrix clearly has $\mathrm{det}(O_p)=-1$ and, when multiplied by all the
elements of $SO(N)$, generates the $\mathrm{det}(V)=-1$ sector of $O(N)$.
So we will use this specific matrix below, with no loss of generality. Now 
under the global gauge transformation $V(x)=O_p$ the $SO(N)$ matrix on
the link $l$ transforms as $U_l\to \tilde{U}_l=O_pU_lO_p$, using
the fact that $O_p^T=O_p$. Since a product of $O(N)$ matrices is in $O(N)$
and since $\mathrm{det}\tilde{U}_l = \{\mathrm{det}O_p\}^2\mathrm{det}U_l = 1$,
the transformed
matrix $\tilde{U}_l$ is in $SO(N)$. Moreover it is clear that the trace of
the ordered product of links around a plaquette is unchanged under this
transformation as is the Haar integration measure since
$d(\tilde{U}_l)=d(\{{U}_l\tilde{U}_l^{-1}\}\tilde{U}_l)=d({U}_l)$.
Thus this transformation is a symmetry of the partition function. If we
now perform some simple algebra, we see that the transformed matrix
$\tilde{U}_l$ is identical to the original matrix ${U}_l$ except that
the first row is multiplied by $-1$ and so is the first column. Note
that this means that the  $(1,1)$ element is multiplied by $(-1)^2=+1$.
This of course applies to any $SO(N)$ matrix. In particular it applies to
a matrix $U_{\mathcal{C}}$ obtained by multiplying the link matrices
around a closed path ${\mathcal{C}}$. It is now trivial to see that
\begin{equation}
  \mathrm{Pf}\{\tilde{U}_{\mathcal{C}}\} = - \mathrm{Pf}\{{U}_{\mathcal{C}}\}.
  \label{eqn_Ultrans}
\end{equation}
We note that this follows from the property of the Pfaffian $\mathrm{Pf}\{A\}$
in eqn(\ref{eqn_PfA}) that if a particular matrix element $A_{ij}$ appears
in a non-zero term of $\mathrm{Pf}\{A\}$, then any other element $A_{kl}$
in that term must have $k\neq i,j$ and $l\neq i,j$, i.e. we have
one (non-diagonal) element from the first row or one from the first column, 
but not from both. Note that the argument will also apply to our generalised
Pfaffians in eqn(\ref{eqn_PfPhiCl}). We can now complete the argument by
noting that the trace is unchanged by this transformation, so
the symmetry implies that
\begin{equation}
  \langle \mathrm{Tr}\{{U}_{\mathcal{C}^\prime}\}\mathrm{Pf}\{{U}_{\mathcal{C}}\} \rangle
=
\langle \mathrm{Tr}\{\tilde{U}_{\mathcal{C}^\prime}\}\mathrm{Pf}\{\tilde{U}_{\mathcal{C}}\}\rangle
=
 -\langle \mathrm{Tr}\{{U}_{\mathcal{C}^\prime}\}\mathrm{Pf}\{{U}_{\mathcal{C}}\} \rangle
= 0
  \label{eqn_PfoverTr}
\end{equation}
for any paths ${\mathcal{C}^\prime}$ and ${\mathcal{C}}$, confirming our
more limited numerical demonstration above. Thus the spectrum
of the theory will contain two sectors, one obtained from correlators
of traces of closed loops and the other from correlators of Pfaffians
of closed loops, with the latter containing our `Pfaffian' particles.

Tha above argument assumes that in our `zero' temperature $SO(N)$
gauge theory this $Z_2$ symmetry is explicit and not spontaneously
broken. Everything we calculate is consistent with that being the case.
A good way to do better would be to identify where the symmetry
is broken (possibly for small spatial volumes or at high temperature)
so as to identify useful lattice order parameters for the symmetry and
then to identify the phase transition where the symmetry is restored.
However such a dedicated investigation lies outside the scope of the present
paper.

\subsection{some other properties}
\label{subsection_further}

We recall that the cyclic property of the trace, i.e.
$\mathrm{Tr}\{\Phi_1\Phi_2\}=\mathrm{Tr}\{\Phi_2\Phi_1\}$,
means that the value of $\mathrm{Tr}\{\Phi(\mathcal{C},n)\}$ will be independent
of the site $n$ along $\mathcal{C}$ at which we choose to take the trace.
We have verified numerically, for simple loops, that for the
$SO(2k)$ gauge theories considered in this paper the same is
true for $\mathrm{Pf}\{\Phi(\mathcal{C},n)\}$: its value is
independent of choice of site $n$ along $\mathcal{C}$.

\section{Pfaffian particles in SO(2N) gauge theories}
\label{section_spectrum}

In this section we will compare the light particle masses as calculated
using the standard trace operators with those obtained using the Pfaffian
operators intoduced above. We compare these spectra with those of the
corresponding unitary gauge theories in those cases where the unitary
and $SO(2k)$ theories share the same Lie algebra, i.e. $SO(2)$ and $U(1)$, 
$SO(4)$ and $SU(2)\times SU(2)$, $SO(6)$ and $SU(4)$. We also calculate
the two types of particle spectra in $SO(8)$ and $SO(10)$ gauge theories so
as to be able to say something about their dependence on $k$ as $k\to\infty$.

The single trace operators whose correlators we calculate are linear combinations
of the traces of our basic loop operators. The linear combinations fall into
separate subsets which are chosen
so that they have specific $J^{P}$ quantum numbers (albeit with the spin ambiguity
described in Section~\ref{subsection_energies}). The single Pfaffian operators
whose correlators we calculate are the same linear combinations of
the Pfaffians rather than the traces of these basic loop
operators. These should have the same spin $J$ as the trace operator but
not necessarily the same parity, as we shall see below. In tabulating
our results for $SO(8)$ and $SO(10)$ we choose to label the `trace' and
`Pfaffian' particles obtained with a given subset of operators by the $J^{P}$
quantum numbers possessed by the `trace' particles. For smaller $k$ where we 
also compare to a unitary theory we display the Pfaffian particles with
different quantum numbers as appropriate.

\subsection{SO(2) and U(1)}
\label{subsection_so2}

The $SO(2)$ gauge theory should have the same physics as the $U(1)$ gauge theory
even if this physics is of limited physical interest. We recall that in units of
the energy scale provided by $g^2$ the $U(1)$ lattice gauge theory is a free
field theory in the continuum limit $\beta=2/ag^2 \to \infty$, but that
at finite $a$ the vacuum contains a screened dilute gas of monopole-like
instantons which lead to a non-trivial mass spectrum and to a non-zero
confining string tension
\cite{U1_conf_a,U1_conf_b,U1_conf_c,U1_conf_d,U1_conf_e}.
These masses vanish exponentially in $\beta=2/ag^2$ as $a\to 0$, reflecting the
similar behaviour of the instanton density, since the instantons are
singular Dirac monopoles.

We can write a general $SO(2)$ matrix assigned to the forward going link $l$ as
\begin{equation}
O_l = \left(
\begin{array}{cc}  
\cos\theta_l & \sin\theta_l \\
- \sin\theta_l & \cos\theta_l
\end{array}  
\right)
  \label{eqn_OlSO2}
\end{equation}
where $\theta_l\in (-\pi,+\pi]$. As usual we assign $O_l^\dagger$ when the link 
is backward going in a path-ordered product.

If we multiply the gauge variables around the closed path $\mathcal{C}$, we obtain some
$SO(2)$ matrix, $O_{\mathcal{C}}$, and hence its gauge-invariant trace and Pfaffian,
\begin{equation}
O_{\mathcal{C}} = \left(
\begin{array}{cc}  
\cos\theta_{\mathcal{C}} & \sin\theta_{\mathcal{C}} \\
- \sin\theta_{\mathcal{C}} & \cos\theta_{\mathcal{C}}
\end{array}  
\right)
\quad ; \quad \mathrm{Tr}\{O_{\mathcal{C}}\} = 2\, \cos\theta_{\mathcal{C}}
\quad ; \quad \mathrm{Pf}\{O_{\mathcal{C}}\} = 2\, \sin\theta_{\mathcal{C}}
  \label{eqn_OcSO2}
\end{equation}
The corresponding $U(1)$ theory has variables on the links that are complex phases
$U_{\mathcal{C}}$:
\begin{equation}
U_{\mathcal{C}} = \exp{i\theta_{\mathcal{C}}}
\quad ; \quad \mathrm{Real}\{U_{\mathcal{C}}\} = \cos\theta_{\mathcal{C}}
\quad ; \quad \mathrm{Im}\{U_{\mathcal{C}}\} = \sin\theta_{\mathcal{C}}
  \label{eqn_PhiSO2}
\end{equation}
where $\mathrm{Real}\{U_{\mathcal{C}}\}$ projects onto the $C=+$ states, 
and $\mathrm{Im}\{U_{\mathcal{C}}\}$ projects onto the  $C=-$ states.
(We drop the trace since we have $1\times 1$ matrices in $U(1)$.) 
In $SO(2)$ we use the  standard plaquette action 
$\beta S[O] = \beta\sum_p \{1-\frac{1}{2}\mathrm{Tr}\{O_P\}\} = \beta\sum_p \{1-\cos\theta_p\}$
where $O_p$ is the product of link matrices around the plaquette $p$,
and in $U(1)$ we use the analogous action action
$\beta S[O] = \beta\sum_p \{1-\mathrm{Real}\{U_P\}\} = \beta\sum_p \{1-\cos\theta_p\}$
where $U_p$ is the product of link matrices around the plaquette $p$.
Since the Haar measure is the same for the two theories, the theories
are identical as indicated by our use of common angular variables. Now
this $SO(2)$ action is invariant under $\theta_l \to -\theta_l \, \forall l$,
so we immediately see that 
\begin{equation}
\langle \mathrm{Tr}O^T_{\mathcal{C}_1} \mathrm{Pf}O_{\mathcal{C}_2} \rangle = 
\langle \mathrm{Tr}O_{\mathcal{C}_1} \mathrm{Pf}O_{\mathcal{C}_2} \rangle = 0 
\quad  \forall{\mathcal{C}_1},{\mathcal{C}_2}.
  \label{eqn_TrPfSO2}
\end{equation}
and similarly if we replace the single trace by a multiple trace operator.
Thus we see that in $SO(2)$ the Trace and Pfaffian project onto two separate
sectors of states, and that these correspond to the $C=+$ and $C=-$ sectors of
the $U(1)$ theory respectively. (And the fact that the product of 2 Pfaffians has
a non-zero overlap onto a traced operator accords with the $C=\pm$ correspondence.)
That is to say, if as in
\cite{RLMT_SON_a,RLMT_SON_b}
we use only single or multiple trace operators we will not be
aware of the existence of a sector of states that is identical to the
$C=-$ sector of $U(1)$. 

It is interesting to see how well one can confirm all the above with an explicit
numerical calculation.
We have therefore calculated the energies of the lightest glueballs
and the lightest flux tube that winds once around one of the periodic spatial
directions. To make the comparison direct we do so on identical $28^236$ lattice
sizes at an identical coupling, $\beta=2.2$, for both $U(1)$ and $SO(2)$.
We first check numerically that, within errors, the states created by the
Pfaffian and Trace operators in $SO(2)$ are indeed orthogonal (as shown
analytically above). Our results for the masses of
the various states are listed in Table~\ref{table_Mso2u1}. We have placed the
Pfaffian results in the row that corresponds to the $J^{PC}$ that one expects
from the fact that the Pfaffian picks out the skew-symmetric piece of
the matrix operator. We see a very nice and convincing match between the masses
of the $SO(2)$ Pfaffian particles and the $C=-$ particles of $U(1)$, at
least for the lightest glueballs where the errors are small. (And of course
the results using traced operators in $SO(2)$ agree with the $C=+$ particles
of $U(1)$.) We also observe that the flux tube energies match very well.
Although we only show one value for $U(1)$ there are in fact two degenerate
ground states of a flux tube. That is to say if we take $l_x$ to be an
operator that winds once around (say) the $x$-direction, then  $l_x^\dagger$
projects onto a flux in the opposite direction. Because of the standard centre
symmatry argument (which for $U(1)$ is the whole group) we know that
$\langle (l_x^\dagger)^\dagger l_x \rangle = 0$ so that these states are orthogonal
and degenerate. So we can choose to use the $l_x \pm l_x^\dagger$
basis, which corresponds to $C=\pm$ respectively and which are clearly
orthogonal and degenerate. It is these that correspond
to the Trace and Pfaffian flux tubes of $SO(2)$ listed in Table~\ref{table_Mso2u1}.
Of course this numerical demonstration that the Pfaffian particles of $SO(2)$
correspond to the $C=-$ states of $U(1)$ is trivial given our earlier discussion.
However it provides a check on the reliability of our numerical calculations,
which is useful for the larger groups considered below.

\subsection{SO(4) and SU(2)$\times$SU(2)}
\label{subsection_so4}

The Lie algebra of $SO(4)$ is the same as that of $SU(2) \times SU(2)$
and so if the different global structures of the groups are unimportant, we
would expect the single particle masses of $SO(4)$ to be the same as those
of $SU(2)$. That this is so has been confirmed, at least for the lightest
masses which are under reasonable control, in the calculations of 
\cite{RLMT_SON_a,RLMT_SON_b,MT_SON}
which used correlators of single trace operators. The present case differs
from that of $SO(2)$ and $U(1)$ discussed above in that $SU(2)$ is
(pseudo)real and therefore there are no $C=-$ states in the correspondence.
However while the spectrum of $SU(2)\times SU(2)$ should contain exactly
the same particle masses as $SU(2)$, these should be doubled, with one from
each of the two $SU(2)$ groups, and this is something that was not
observed in
\cite{RLMT_SON_a,RLMT_SON_b,MT_SON}
although this fact was not remarked upon in those papers. On the other hand
the $SO(4)$ theory should contain only one vacuum state just like
$SU(2)\times SU(2)$: that state should
not be doubled. And the lightest flux tube, carrying the fundamental flux
of $SO(4)$, should also not be doubled. As we shall now see,
including Pfaffian operators will in fact allow us to meet all
these expectations.

In this calculation we shall not attempt to perform continuum extrapolations 
of the masses as in 
\cite{RLMT_SON_a,RLMT_SON_b,MT_SON}
but rather we choose a single coupling at which the spectrum is very close to
its continuum limit. For our $SO(4)$ calculation we choose to use  $\beta=15.1$
on a $50^256$ lattice, which we expect to have negligible finite volume
corrections for the masses calculated. We then use the $SU(2)$ calculations of
\cite{AAMT_SUN}
to choose a value $\beta=13.87$, where we expect the mass gap to equal that
of the $SO(4)$ calculation within statistical errors. The results of these
calculations for the lightest states are listed in Table~\ref{table_Mso4su2}.
We see that the very lightest Pfaffian particles listed are consistent with
being degenerate with the corresponding `trace' particles and with the
corresponding $SU(2)$ masses, within statistical errors. Heavier mass
estimates will be afflicted by increasing systematic errors
and so are less reliable. Thus we see that the Pfaffian particles
indeed appear to provide us with the expected doubling in the $SO(4)$
spectrum.

\begin{figure}[htb]
\begin	{center}
\leavevmode
\input	{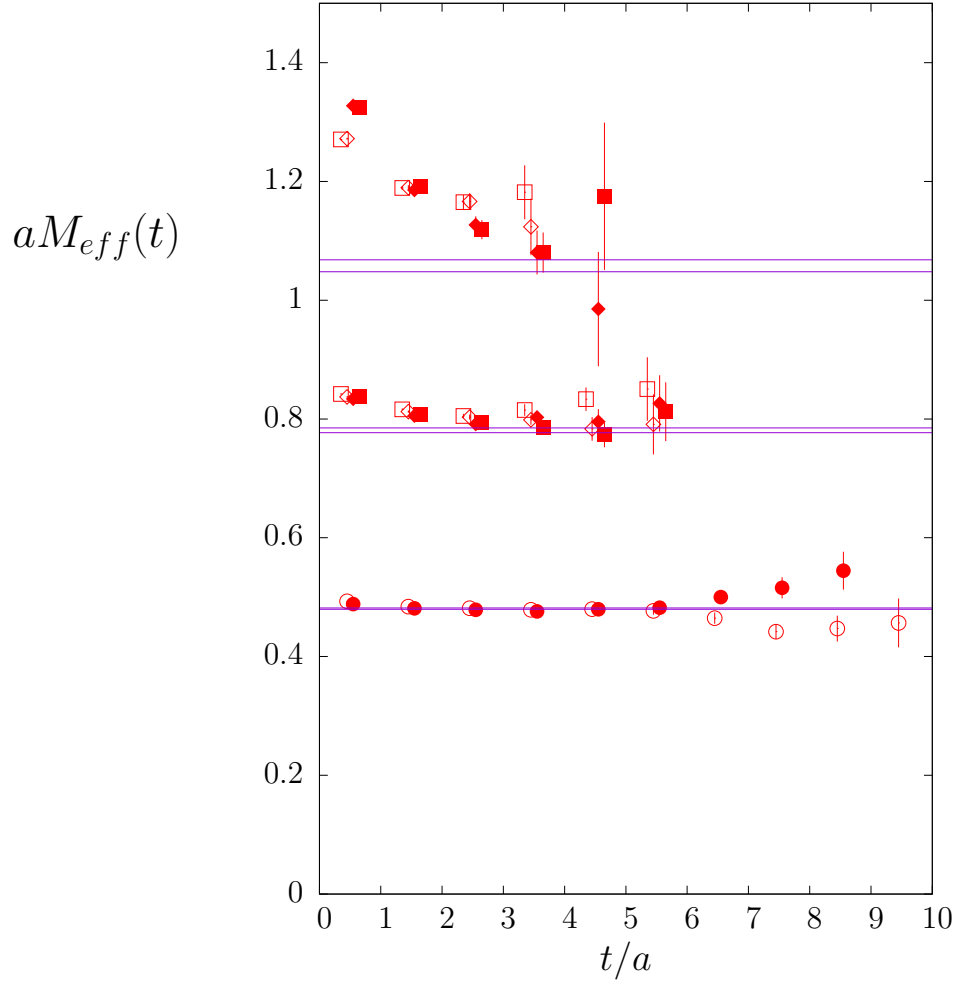}
\end	{center}
\caption{Effective energies of the lightest glueballs in $SO(4)$ at $\beta=15.1$
  on a $50^2 56$ lattice: trace $J^P=0^{+}$, $\circ$,  Pfaffian `$0^{+}$',$\bullet$,
  trace $2^{\pm}$, lower $\lozenge, \square$, Pfaffian
  $2^{\pm}$, lower $\blacklozenge, \blacksquare$
  trace $1^{\pm}$, higher $\lozenge, \square$, Pfaffian
  $1^{\pm}$, higher $\blacklozenge, \blacksquare$. Lines
  are  $0^{+}$, $2^{\pm}$, $1^{\pm}$ masses (in ascending order) obtained
  in $SU(2)$ at $\beta=13.87$ on the same size lattice. Points shifted
  for clarity.}
\label{fig_EgeffPfTrso4}
\end{figure}

It will be useful for the reader to see examples of the effective mass
plots that are behind our mass estimates. In Fig.~\ref{fig_EgeffPfTrso4}
we show plots of $aM_{eff}(t)$ against $t$ for the lightest $J^P=0^+$,
$J^P=2^\pm$, and  $J^P=1^\pm$ states obtained with trace and Pfaffian
operators. We simultaneously show the asymptotic mass estimates for the
same states in the $SU(2)$ theory. We see that the quality of the
$0^+$ comparison is excellent, that for the $2^\pm$ it is quite convincing,
while for the $1^\pm$ it is indicative but the lack of a clear effective
mass plateau means that it is not much more than that. (Although we note
that the evidence from the plot that the $J=1$ trace mass is
degenerate with the $J=1$ Pfaffian  mass is much more convincing.)
This illustrates the main issue with extracting heavier masses. We also
note that some effective masses appear to increase at  larger $t$.
By the positivity of the correlator we know that this is not possible,
and so this must be a statistical fluctuation even though the errors shown
might suggest otherwise. This is confirmed by the fact that the $2^+$
and $2^-$ should be degenerate. In any case, despite the caveats, this
plot does illustrate quite clearly the degeneracy of the trace and
Pfaffian masses.

The above comparison is on a lattice that is large enough for finite
volume corrections to be invisible for the states we consider.
One can also ask if the $SO(4)$ and $SU(2)$ particle masses remain
the same when calculated on smaller volumes where some of the states are
affected by finite volume corrections. So in Table~\ref{table_Mso4su2_small}
we show the results of a calculation on a smaller $34^2 56$ lattice
at the same couplings as above. The significant breaking of the $2^\pm$
degeneracy in $SU(2)$ is a finite volume effect
\cite{AAMT_SUN}.
We see that the $SO(4)$ Pfaffian spectrum exhibits a similar finite volume
breaking of the  $2^\pm$ degeneracy and that it provides an acceptable
match to both the $SO(4)$ trace spectrum and to the $SU(2)$ spectrum.

Numerically we find that the Pfaffian and trace states appear to be orthogonal.
This manifests itself in a striking way in the fact that the Pfaffian $0^+$
operators have zero overlap onto the vacuum while the corresponding trace
operators have a very large vacuum overlap, which is subtracted in the
calculation so as to expose the $0^+$ ground state glueball. That is to
say, the union of trace and Pfaffian operators does indeed produce pairs of
degenerate glueballs but only a single vacuum state.

\subsection{SO(6) and SU(4)}
\label{subsection_so6}

Earlier calculations of the light glueball spectrum in the $SO(6)$ gauge theory
\cite{RLMT_SON_a,RLMT_SON_b,MT_SON}
have provided strong evidence that the part of the spectrum that can be obtained
from correlators of single trace operators coincides with the $C=+$ spectrum
of the $SU(4)$ gauge theory. As remarked in the Introduction, it is hard
to understand how the heavier  $C=+$ glueballs, which in the $SU(4)$ theory
can decay into a pair of lighter $C=-$ glueballs, can be the same in both
$SO(6)$ and $SU(4)$ -- including their decay width -- unless the $SO(6)$ gauge
theory possesses particles degenerate with the  $C=-$ glueballs of  $SU(4)$.

As we have just seen above, in $SO(2)$ the Pfaffian particles are precisely the
$C=-$ particles of $U(1)$. Both $SO(2)$ and $SO(6)$ belong to the $SO(4k+2)$
series within which, we argued earlier, the Pfaffian operators are orthogonal
to the trace operators, something that we have confirmed numerically to be the
case (within small errors). Thus it is natural to conjecture that the Pfaffian
particles of $SO(6)$ correspond to the $C=-$ particles of $SU(4)$.

To test this conjecture we have performed calculations in $SO(6)$ and $SU(4)$
at lattice bare couplings $\beta=46.0$ and $\beta=59.14$ respectively, which
have been chosen so as to lead to the same mass gap (within errors) in the
two theories, and also to be close to the continuum limit (for the
quantities previously calculated). We use the same $46^2 48$ lattice in
both cases. The resulting masses are listed in Table~\ref{table_Mso6su4}.
The Pfaffian particles have been placed in the same rows as in our
earlier $SO(2)$ calculations. We observe a convincing confirmation of our
conjecture that the Pfaffian particles of $SO(6)$ correspond to the $C=-$
particles of the $SU(4)$ gauge theory, at least for the lightest states
where the error estimates are under reasonable control.
In Table~\ref{table_Mso6su4_small} we repeat the exercise on a smaller
$36^2 44$ lattice, with the same conclusion. All this provides strong
numerical evidence that the spectra of the $SU(4)$ and $SO(6)$ gauge theories
are indeed the same, with the Pfaffian particles of the latter
corresponding to the $C=-$ particles of the former. 

\begin{figure}[htb]
\begin	{center}
\leavevmode
\input	{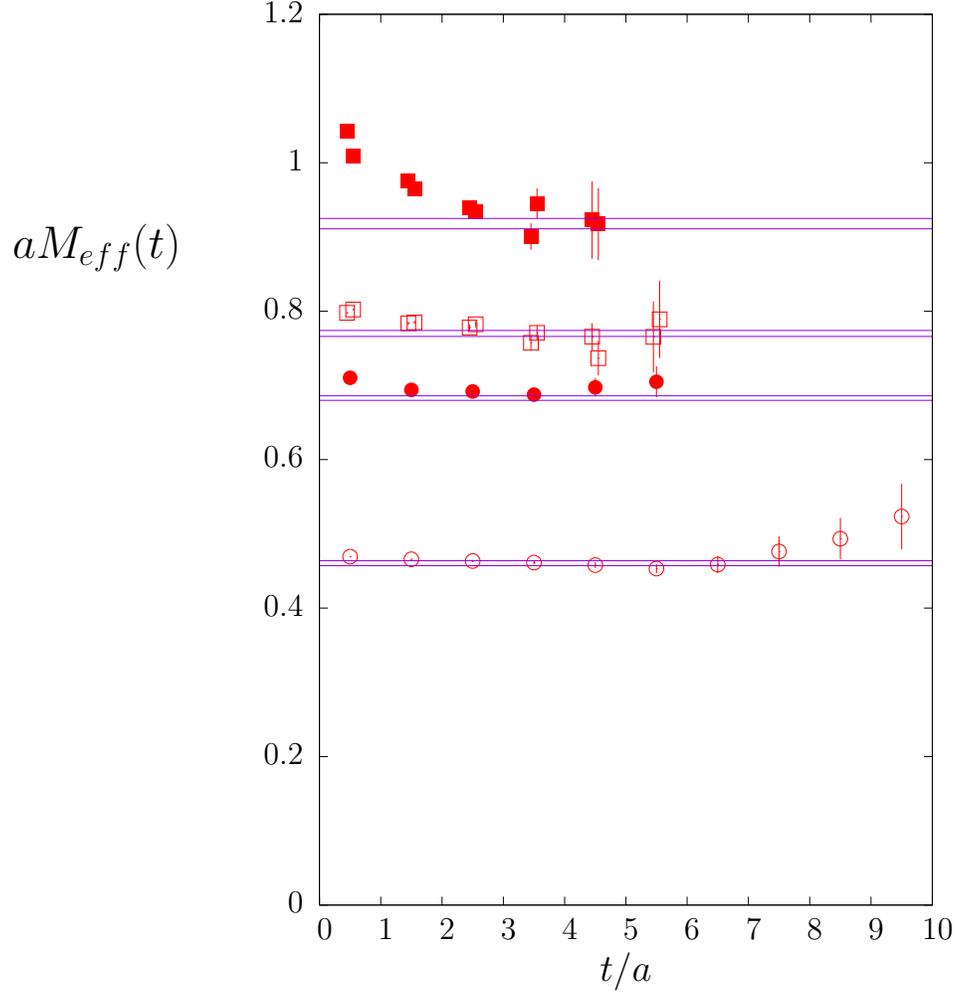}
\end	{center}
\caption{Effective energies of the lightest glueballs in $SO(6)$ at $\beta=46.0$
  on a $46^2 48$ lattice: trace $J^P=0^{+}$, $\circ$ and trace $2^{\pm}$, $\square$
  with Pfaffian `$0^{+}$',$\bullet$, and Pfaffian `$2^{\pm}$', $\blacksquare$. Pairs
  of lines are  $\pm 1\sigma$ bands for the $J^{PC}=0^{++}$, $0^{--}$, $2^{\pm +}$, $2^{\pm -}$ masses (in ascending order) obtained in $SU(4)$ at $\beta=59.14$ on the same size lattice.}
\label{fig_EgeffPfTrso6}
\end{figure}

As in the case of $SO(4)$, it is useful to display some of the effective
mass plots that underpin our $SO(6)$ mass estimates. This we do in
Fig.~\ref{fig_EgeffPfTrso6} for the lightest trace and Pfaffian `$0^{+}$'
and `$2^{\pm}$' states. We see reasonably identifiable effective mass
plateaux in all cases. (Just as in $SO(4)$ this is not so clear for
the $1^\pm$ states, which are significantly heavier than those shown.)
In contrast to $SO(4)$ it is clear that the lightest trace and
Pfaffian particles have very different masses. We also show the
asymptotic mass estimates in for the lightest
$0^{++},0^{--},2^{\pm +},2^{\pm -}$ particles in $SU(4)$. It is quite
clear from this that while the trace particles correspond to the
$C=+$ particles of $SU(4)$, the Pfaffian particles correspond to the
$C=-$ particles, as conjectured above.

\subsection{SO(8) and SO(10)}
\label{subsection_so8so10}

For $SO(2k > 6)$ there is no $SU(N)$ group with the same Lie algebra. Since
nonetheless the perturbative planar $N\to\infty$ limits of $SU(N)$ and $SO(N)$
gauge theories coincide
\cite{largeN_SON}
it is natural to conjecture that the ($C=+$) glueball spectra will also
coincide. And there is strong numerical evidence
\cite{RLMT_SON_a,RLMT_SON_b,MT_SON}
that this is indeed so. To get numerical evidence for the fate of
Pfaffian particles in this limit we need some calculations for $2k > 6$
and so we have performed calculations in $SO(8)$ and in $SO(10)$.

Our calculations in $SO(8)$ are at $\beta=84.0$ on a $28^236$ lattice, and
the resulting glueball mass estimates are listed in Table~\ref{table_Mso8so10}.
In units of the inverse mass gap the spatial size is $laM_{0^{++}} \sim 14$
which is a little smaller than the smaller of the two lattices we used for
$SO(6)$. To check for finite volume corrections we have also performed
calculations on a $22^240$ lattice at the same coupling corresponding to the
smaller spatial size  $laM_{0^{++}} \sim 11$. We find that the masses are
the same within errors so we assume that any finite volume effects are
negligible for our purposes.

In Table~\ref{table_Mso8so10} we also list the results of our glueball mass
calculation in $SO(10)$. This is on a $22^230$ lattice at $\beta=120.0$.
Here $laM_{0^{++}} \sim 13$ which we shall assume has no significant finite
volume corrections given what we have just observed for $SO(8)$.

\subsection{N dependence}
\label{subsection_largeN}

Since the $SO(2k)$ Pfaffian involves a product of $k$ fields, it is natural
to expect the Pfaffian particle masses to increase $\propto k$ as $k$ increases,
just like baryons in $N$-colour QCD. It is interesting to test this
expectation and also to see whether the sub-leading corrections to this
behaviour are small, just as they have been found to be for the `trace'
particles
\cite{RLMT_SON_a,RLMT_SON_b,MT_SON}.

\begin{figure}[htb]
\begin	{center}
\leavevmode
\input	{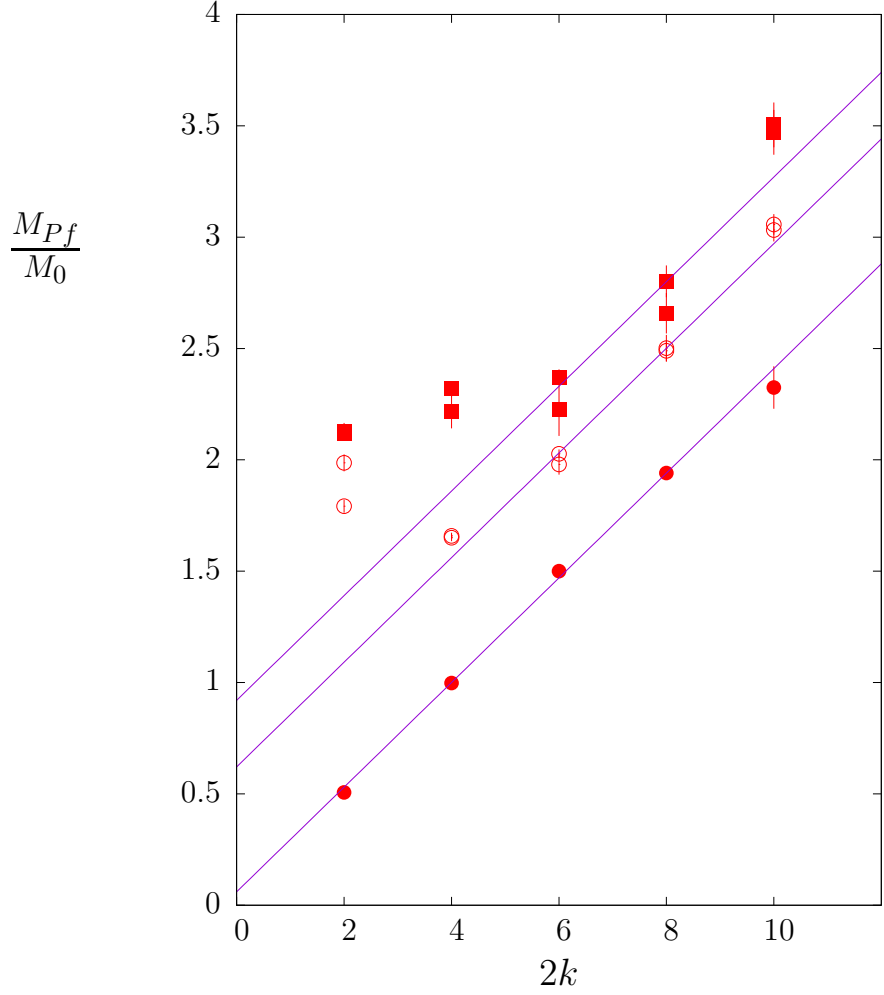}
\end	{center}
\caption{Lightest Pfaffian masses in units of the lightest trace $J^{P}=0^{+}$
  mass in our various $SO(2k)$ gauge theories:
  $J=0$ ($\bullet$), $J^P=2^\pm$ ($\circ$) and $J^P=1^\pm$ ($\blacksquare$).}
\label{fig_MPfM0}
\end{figure}

In Fig.~\ref{fig_MPfM0} we display the masses of the lightest
$J=0$, $J=2$ and $J=1$ Pfaffian particles in units of the the lightest
trace $J^{P}=0^{+}$ glueball for our $SO(2k)$ groups. (Masses taken from
Tables~\ref{table_Mso2u1},\ref{table_Mso4su2},\ref{table_Mso6su4},\ref{table_Mso8so10}.)
The most striking feature of this plot is the nearly linear growth with $k$
of the mass of the $J=0$ Pfaffian glueball: the fit in the plot is
simply $M_{J=0}^{Pf}/M_{0^{+}} = 0.06 + 0.47k$. This is the lightest and hence 
most accurately measured of our masses. We use a subleading correction
that is down by a single power of $k$ because in $SO(N)$ gauge theories
this is the case in general when one considers diagrams
\cite{largeN_SON}.
(Caveat: we have not shown that this is the case for Pfaffian
operators but we expect it to be so.)
We show similar linear fits for $J=2$ and $J=1$ Pfaffian particles,
where we have constrained the coefficient of the linear piece to be
independent of $J$. For $J=2$ this appears to work well except for the
very lowest value of $k$, while for the heavier $J=1$ Pfaffian particles
one has to go to larger $k$ to see this linear rise. This is no surprise:
as the particles become heavier the subleading correction becomes larger
and presumably so do the higher order corrections in $k$, and this will provide
an increasing curvature to the dependence of the masses on $k$.
In summary we see from  Fig.~\ref{fig_MPfM0} clear evidence for the
asymptotic linear growth with $k$ for the Pfaffian particles.

The fact that the  lightest Pfaffian particle has small corrections
to the leading $\propto k$ behaviour encourages us to make the
following simple argument for the mass of the lightest $J=0$ Pfaffian
particle in units of the lightest (trace) $J^P=0^{++}$ mass i.e. for
the ratio $m_0^{Pf}/m_{0^{++}}$. For $U(1)$ we expect the continuum theory
to be a free theory of $J^{PC}=0^{--}$ particles. The lightest $0^{++}$
state is composed of two non-interacting $0^{--}$ particles so we expect
$m_0^{Pf}/m_{0^{++}} = m_{0^{--}}/m_{0^{++}} = 2$ using the identity
between $SO(2)$ and $U(1)$. For $SO(4)$ we expect, as argued above, that 
$m_0^{Pf}/m_{0^{++}} = 1$. Assuming no corrections to the
expected large-$k$ behaviour of $m_0^{Pf}/m_{0^{++}} \propto k$ we
infer that in $SO(6)$ we have $m_0^{Pf}/m_{0^{++}} = 1.5$. Since the
ground state $J=0$ Pfaffian corresponds to the $0^{--}$ of $SU(4)$
we have the prediction for $SU(4)$ that  $m_{0^{--}}/m_{0^{++}} = 1.5$
which provides a good approximation to the calculated value
\cite{AAMT_SUN}
of $m_{0^{--}}/m_{0^{++}} = 1.465(5)$. Of course this is no `vanilla'
prediction: we need to use the fact that the observed corrections
to the leading $k$ dependence are small.

\section{Pfaffian strings and confinement}
\label{section_confinement}

In $SU(N)$ gauge theories there is the well-known and elegant connection between the
(spontaneous breaking of the) $Z_N$ centre symmetry of $SU(N)$ and (de)confinement.
As reviewed in Section~\ref{subsection_stringPfaffian}
this same argument extends to the $Z_2$ centre of $SO(2k)$: in the confining
phase a flux loop $\Phi_l$ that winds once around the periodic $x$ direction will
satisfy $\langle \mathrm{Tr} \{\Phi_l\} \rangle = 0$ and
$\langle \mathrm{Tr} \{\Phi^\dagger_c \}\mathrm{Tr} \{\Phi_l \}\rangle = 0$
where  $\Phi_c$ is any contractible loop. The finite volume states to which
$\mathrm{Tr} \{\Phi_l \}\rangle$
couples are flux tubes that wind around the $x$-torus. As shown in
\cite{RLMT_SON_a,RLMT_SON_b}
the resulting string tensions are consistent between $SU(4)$ and $SO(6)$
if one takes into account the fact that the fundamental $SO(6)$ flux
corresponds to the totally antisymmetric piece of $f\otimes f$ in $SU(4)$,
usually labelled as $k=2A$. They are also consistent between $SO(4)$ and twice
the string tension of $SU(2)$ (twice because of the two $SU(2)$ groups)
and, indeed, between $SO(N\to\infty)$ and  $SU(N\to\infty)$ when expressed,
for example, in units of the mass gap in each theory
\cite{RLMT_SON_a,RLMT_SON_b}.

All this is for trace operators. What happens when we take the Pfaffian
of $\Phi_l$ instead of the trace? As remarked earlier, under the $Z_2$
of  $SO(2k)$ we have $\Phi_l \to -\Phi_l$
and hence $\mathrm{Pf}\{\Phi_l\} \to (-1)^k \mathrm{Pf}\{\Phi_l\}$ since
$\mathrm{Pf}\{\Phi_l\}$
contains a product of (pieces of) $\Phi_l$ $k$ times. Thus it is natural to expect
that in $SO(4k+2)$ $\mathrm{Pf}\{\Phi_l\}$ will project onto some kind of confining
flux tube wrapped around the $x$-torus, but not necessarily so in  $SO(4k)$.
That is to say, the energy of the lightest state obtained from correlators
of $\mathrm{Pf}\{\Phi_l\}$ should grow roughly linearly with the length $l_x$
in the case of $SO(4k+2)$, but maybe not in $SO(4k)$. We shall now try to
determine what actually happens in the gauge groups investigated in this paper.
We begin with $SO(2)$ and $SO(6)$ which belong to the $SO(4k+2)$ series
and then move on to  $SO(4)$ which belongs to the $SO(4k)$ series. Finally
we briefy consider larger $k$.

\subsection{SO(2) and U(1)}
\label{subsection_string_so2}

We showed earlier that the trace and Pfaffian in $SO(2)$
correspond to the real and imaginary parts of the trace in $U(1)$.
In $U(1)$ the real and imaginary parts of a flux loop operator are simply
$2\mathrm{ReTr} \{\Phi_l\} =  \mathrm{Tr}\{\Phi_l\} +  \mathrm{Tr}\{\Phi_l^\dagger\}$ and
$2\mathrm{ImTr} \{\Phi_l\} =  \mathrm{Tr}\{\Phi_l\} -  \mathrm{Tr}\{\Phi_l^\dagger$\} and
these are $C=+$ and $C=-$ flux loops respectively. The usual centre symmetry
argument ensures that the correlator of $\mathrm{Tr}\{\Phi_l\}$ with 
$\mathrm{Tr}\{\Phi_l^\dagger\}$ is zero, which means that the  $C=+$ and $C=-$
correlators are identical and hence that the $C=+$ and $C=-$ flux loop
spectra are degenerate. The same argument will hold for $SU(N>2)$ which
is why in all these cases one normally quotes a single string tension,
although strictly speaking there are two equal ones corresponding
to  $C=+$ and $C=-$. With $SO(2)$ the $C=+$ and $C=-$ flux loops
correspond to trace and Pfaffian operators respectively, which are
mutually orthogonal, and produce equal string tensions as we see in
Table~\ref{table_KsoN}. As expected we also see in Table~\ref{table_KsoN}
that this string tension equals the corresponding $U(1)$ string tension.
Given the evident identity between $SO(2)$ and $U(1)$ and given the fact
that the latter is well known
\cite{U1_conf_a,U1_conf_b,U1_conf_c,U1_conf_d,U1_conf_e}
to possess linear confinement, we do not show here any numerical results
displaying this fact.

\subsection{SO(6) and SU(4)}
\label{subsection_string_so6}

One expects the fundamental flux of $SO(6)$ to correspond to the
$k=2A$ flux in $SU(4)$. (As usual $k=2A$ denotes the totally
antisymmetric piece of $f\otimes f$.) In
\cite{RLMT_SON_a,RLMT_SON_b}
it was shown that, within small errors, the energy of the lightest
(traced) flux loop is indeed the same as that of the lightest $k=2A$
flux loop in $SU(4)$ and displays the same nearly-linear growth
with length. In the minimal $k=2$ sector of $SU(4)$ there is in addition
to the $k=2A$ representation also the  $k=2S$ representation which
is the totally symmetric piece of $f\otimes f$
\cite{BLMT}.
(There are of course larger  $k=2$ representations obtained, for example,
from $f\otimes f\otimes f\otimes \bar{f}$.) Since the trace in the $k=2A$
representation is real while that of  $k=2S$ contains an imaginary
piece, one might conjecture that the Pfaffian in $SO(6)$ maps to
the  $k=2S$ of $SU(4)$ and that the operator $\mathrm{Pf}\{\Phi_l\}$
projects onto winding flux tubes that correspond to those in
$SU(4)$ carrying $k=2S$ flux. We shall now see that this conjecture
is (largely) correct.

\begin{figure}[htb]
\begin	{center}
\leavevmode
\input	{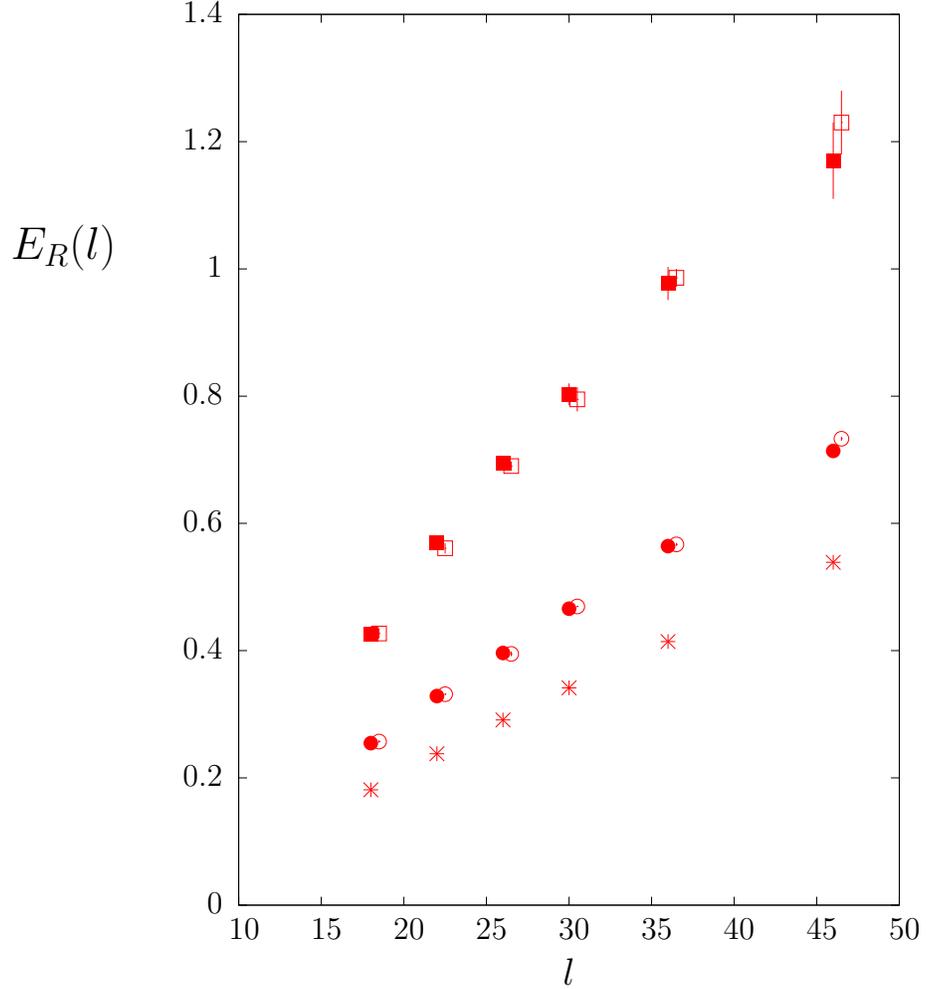}
\end	{center}
\caption{Energy of lightest flux tube against its length: carrying fundamental
  ($\star$), $k=2A$ ($\circ$) and $k=2S$ ($\square$) flux in $SU(4)$ at $\beta=59.14$
  and trace ($\bullet$) and Pfaffian ($\blacksquare$) in $SO(6)$ at $\beta=46.0$.
  Some points shifted slightly for visibility.}
\label{fig_Elsu4so6}
\end{figure}

We have performed calculations in $SO(6)$ at $\beta=46.0$ on
$l\times l_y\times l_t$ lattices and have calculated the spectrum
obtained from correlators of traces and Pfaffians of operators
$\Phi_l$ that wind once around the $x$ direction of length $l$.
As $l$ decreases we increase the transverse size $l_y$ and also $l_t$
so as to minimise finite transverse volume and finite temperature effects.
The lattice sizes are listed in Table~\ref{table_Elso6su4} where we
also list the lightest energies we obtain from our operators of the
form $\mathrm{Pf}\{\Phi_l\}$ and $\mathrm{Tr}\{\Phi_l\}$. These energies are
plotted in Fig~\ref{fig_Elsu4so6} where we see that the Pfaffian
energy grow linearly just like that of the usual (traced) loop.
That is to say, the Pfaffian loop operator does indeed project onto
states that consist of flux tubes winding around the spatial torus,
just like the standard trace operator. It is however orthogonal to the
latter and the string tension (slope) is clearly very different.

Just as for the glueballs it is useful to show the effective energy plots
that are behind this calculation. This we do for the Pfaffian strings
in Fig.~\ref{fig_EleffPfso6} and for the trace strings in 
Fig.~\ref{fig_EleffTrso6}, and we compare them to our asymptotic
energy estimates. It is clear that the trace energies are
accurate and unambiguous. The Pfaffians have worse overlaps
and the energies are larger, all of which makes the calculations
much less reliable. For the Pfaffian strings the identification of
effective mass plateaux is moderately convincing for $l\leq 26$ and
perhaps also for $l=30$, but for $l=36$ and particularly for $l=46$
one has to assume, on the basis of what one sees at smaller $l$,
that one is close to a plateau by $t\sim 3.5a$ in order to extract
any energy at all.

To make the comparison with $SU(4)$ we perform calculations in $SU(4)$
on exactly the same lattice sizes at the coupling $\beta=59.14$
where the mass gap equals the mass gap in $SO(6)$ at $\beta=46.0$.
We calculate the lightest energies of flux loops carrying
fundamental ($k=1$), $k=2A$ and $k=2S$ flux versus the length $l$
of the loop. The results are displayed in Fig~\ref{fig_Elsu4so6}.
We observe that the $k=2A$ energies of $SU(4)$ are indeed
degenerate with those of the traced loops in $SO(6)$, as observed
in  earlier work
\cite{RLMT_SON_a,RLMT_SON_b}.
More interestingly we observe a similar degeneracy between the
$k=2S$ energies of $SU(4)$ and the Pfaffian flux loops in $SO(6)$,
providing strong numerical evidence for our above conjecture.

However this conjecture can only be `largely' correct. The Pfaffian
and trace flux loop operators in $SO(6)$ are orthogonal. This
is not the case for $k=2A$ and $k=2S$ in $SU(4)$ although in practice
they are very nearly orthogonal
\cite{2A2S_over}.
The mixing between $k=2A$ and $k=2S$ flux tubes may be driven by
tunnelling and it may be that this is the kind of physics in which
the groups will differ even if the Lie algebras are the same.
In any case while all this means that our matching between $SO(6)$ and
$SU(4)$ flux loops is entirely adequate for most practical purposes,
the theoretical underpinning is not yet complete.

\subsection{SO(4) and SU(2)$\times$SU(2)}
\label{subsection_string_so4}

Unlike $SO(2)$ and $SO(6)$ the group $SO(4)$ does not belong to the $SO(4k+2)$
series so although we know that the $SO(4)$ theory is linearly confining
\cite{RLMT_SON_a,RLMT_SON_b}
we are not confident that the Pfaffian of a flux loop operator projects onto
some kind of confining flux tube. To investigate this question we have
performed calculations in $SO(4)$ and $SU(2)$ at $\beta=15.1$ and
$\beta=13.87$ respectively,
at which $\beta$ values the  $SO(4)$ and $SU(2)$ mass gaps are equal (within
small errors). We perform calculations for various vaues of $l$ in  $SO(4)$
and for three values for $SU(2)$. (These include the smallest and largest
$SO(4)$ values of $l$, as well as an intermediate value.) 

The expectation is that the string tension in $SO(4)$ is twice that
of $SU(2)$. To compare our $SU(2)$ flux loop energies to those of
$SO(4)$ we proceed as follows. For each value of $l$ we extract the
$SU(2)$ string tension using the formula in eqn(\ref{eqn_NG}). We then
double that string tension and calculate the flux loop with the
doubled string tension, again using eqn(\ref{eqn_NG}).
This energy is listed in Table~\ref{table_Elso4su2}
as $E_{eff}$. We see from the table that these values agree reasonably
well with the energies $E^{Tr}$ of the traced flux loops in $SO(4)$,
confirming the conclusions of earlier work
\cite{RLMT_SON_a,RLMT_SON_b}.

Our calculation of the lightest Pfaffian flux loop turns out to be more
complex. As usual our variational procedure maximises $\exp(-aE)$ and we
calculate the energy of the state from the effective mass plateau
of the corresponding correlator. Normally this gives us the lightest
energy (as can be checked by looking at the correlators of the
higher excited states). However this is not guaranteed: if the lightest
state has a very poor overlap onto our basis of operators
it may appear in the large $t$ tail of a correlator that
one would normally expect to correspond to a heavier excited state.
This is what we find with the Pfaffian of the flux loop.
In  Table~\ref{table_Elso4su2} we list the energy obtained from
operator that maximises $\exp(-aE)$ as $E^{Pf}$ and, where 
lighter states appear in the large $t$ tails of what should be
excited states, we list the lightest of these as $\tilde{E}^{Pf}$.
As $l$ increases the overlap of this lightest state decreases
and so it is harder to identify an effective mass plateau, so
we place the $l=46,50$ values in brackets to indicate this
uncertainty. For $l=18$ the lightest state is indeed the one that
maximises  $\exp(-aE)$ so we show the next energy as $\tilde{E}^{Pf}$
and again place it in brackets to indicate some uncertainty in this
assignment. Even if we ignore the bracketed values of $\tilde{E}^{Pf}$
it appears that the Pfaffian projects onto at least one state
whose energy does not increase with $l$, and which is therefore
particle-like rather than string-like, although its overlap
is very small and appears to decrease with increasing $l$.
In contrast, the lightest state with a substantial overlap onto our
pfaffian basis has an energy $E^{Pf}$ that increases with $l$ as
one would exect for some kind of stringy flux loop.

\begin{figure}[htb]
\begin	{center}
\leavevmode
\input	{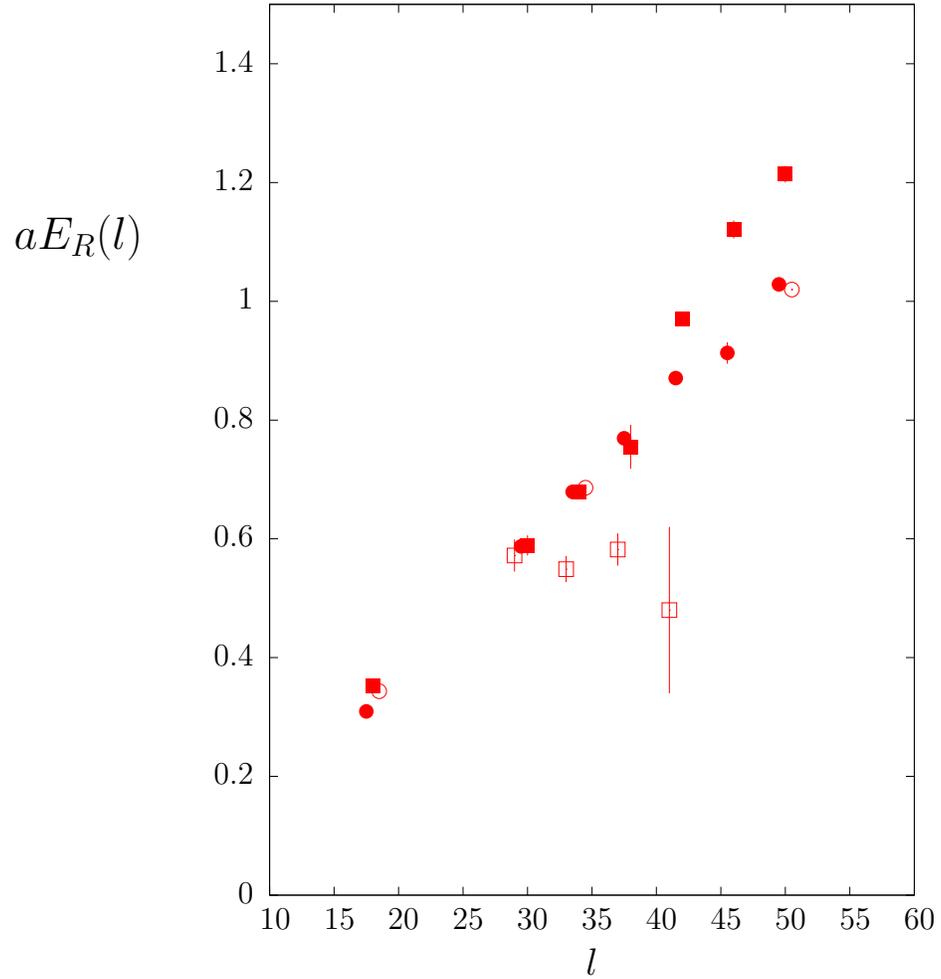}
\end	{center}
\caption{Energy of lightest flux tube against its length: carrying fundamental
  ($\circ$) flux in $SU(2)$ at $\beta=13.87$,
  and trace ($\bullet$) and Pfaffian ($\square,\blacksquare$) in $SO(4)$ at
  $\beta=15.1$. See text for difference between the two Pfaffian entries.
  Some points shifted slightly for visibility.}
\label{fig_Elso4su2}
\end{figure}

To better expose these spectra we plot the energies in
Fig~\ref{fig_Elso4su2}. We observe the nearly linear rise with $l$
of  $E^{Tr}$ and the fact that it is compatible with what one expects
from $SU(2)$. The `normal' Pfaffian energy, $E^{Pf}$, appears to
be nearly degenerate with $E^{Tr}$ at small $l$ but appears to
grow faster at the largest values of $l$. However this latter
behaviour may be illusory: the energies are becoming large, the
overlaps are mediocre and so we may be overestimating the energies
by not going far enough in $t$ to identify the effective energy
plateau. The presence of a particle-like lighter state with energy
$\tilde{E}^{Pf}$ appears to be unambiguous. However a more accurate
calculation is clearly needed here.

\begin{figure}[htb]
\begin	{center}
\leavevmode
\input	{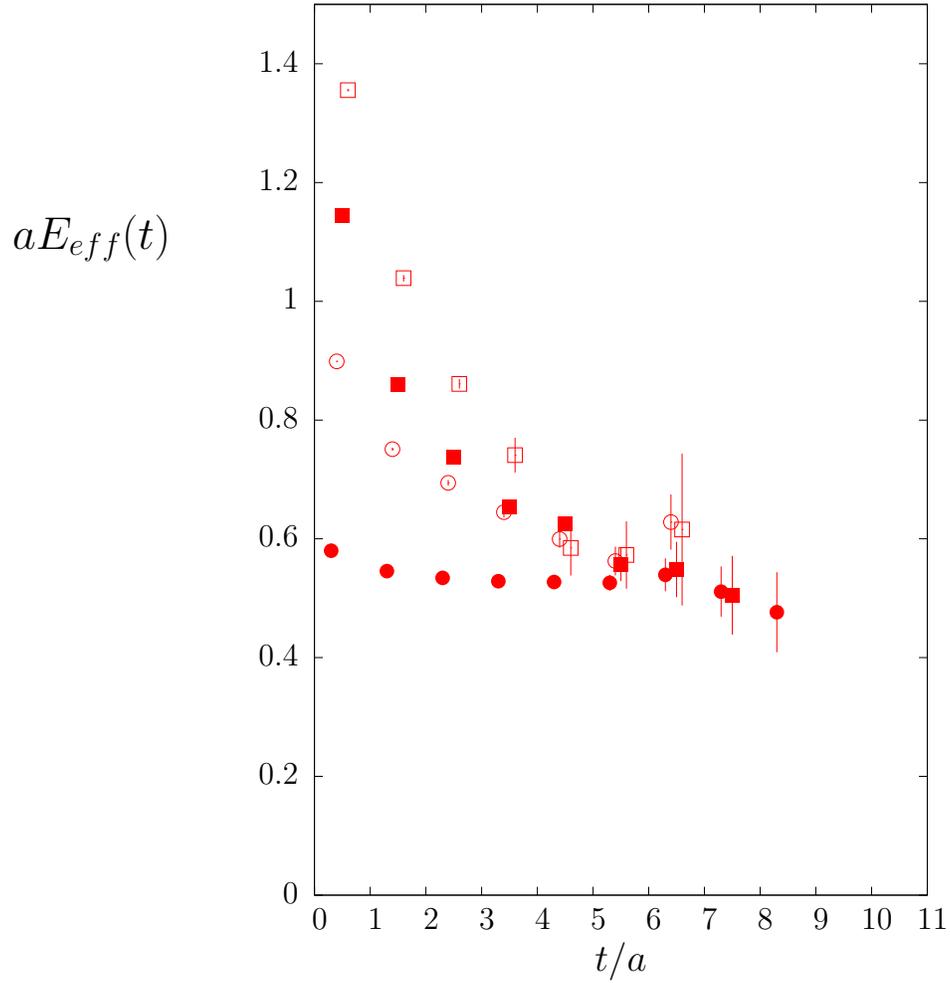}
\end	{center}
\caption{Effective energies of the particle-like state
  in the Pfaffian `string' spectrum in $SO(4)$ at $\beta=15.1$
  for lengths $l=18,\bullet$,$l=30,\circ$,$l=34,\blacksquare$,$l=38,\square$.
  Some points shifted slightly for visibility}
\label{fig_EleffPfexso4}
\end{figure}

To show how reliable are the above observations, we display
the effective energy plots for the traced flux loop in
Fig.~\ref{fig_EleffTrso4}, for the string-like Pfaffian in
Fig.~\ref{fig_EleffPfso4} and for the  particle-like Pfaffian in
Fig.~\ref{fig_EleffPfexso4}. The traced flux loop clearly has
a very good overlap onto our basis of operators and so has convincing
effective energy plateaux, except for $l=46$ which appears to suffer
a large statistical fluctuation which we try to encompass with
larger errors on our final energy estimate. From Fig.~\ref{fig_EleffPfso4}
we see that the overlaps of the Pfaffian string-like states are
much poorer leading to the plateaux being at larger $t$ where
the larger errors make the identification more difficult.
However for the most part our energy estimates are quite plausible,
even if the unquantified systematic errors, which grow with $l$,
leave some room for doubt, particularly at larger $l$.
Finally we see in Fig.~\ref{fig_EleffPfexso4} that the particle-like
Pfaffian has a very poor overlap onto our basis for almost all $l$
which means that identification of an effective energy `plateau'
around $t\sim 6a$ isvery subjective. The exception is $l=18$ where
the overlap is better and we have a decent plateau -- but as we
remarked above, there is some uncertainty in ascribing this
state to belong to the particle-like family. In any case the
positivity of the correlator guarantees that $E_{eff}(t)$ always
provides an upper bound for the true energy, and it is quite
clear from Fig.~\ref{fig_EleffPfexso4} that this energy does
not grow with $l$ and hence this state is not some kind of
confining flux loop, but must be essentially particle-like.

\subsection{larger N}
\label{subsection_string_soN}

We have also performed some calculations of the trace and Pfaffian of
flux loop operators in $SO(8)$ and $SO(10)$. In $SO(10)$ we have
calculated the energies for only one value of $l$ since this belongs
to the $SO(4k+2)$ series where we expect both the trace and the Pfaffian
of flux loop operators to project onto stringy states whose energies grow
roughly linearly with $l$. The energy is listed in Table~\ref{table_KsoN}
together with the string tension extracted using eqn(\ref{eqn_NG}).

For $SO(8)$ we have two values of $l$. The energies and corresponding
string tensions are listed in Table~\ref{table_KsoN}. We see that the
string tensions from the trace are equal within errors and that the Pfaffian
string tensions are compatible with each other. That is to say, both the
trace and Pfaffian of a flux loop project onto states that are stringy
with an energy that grows almost linearly with $l$. The trace
and Pfaffian string
tensions are however very different. In addition although $SO(8)$
falls into the $SO(4k)$ series, just like $SO(4)$, there is no sign
of a lighter particle-like state hidden amongst the excited state
correlators. This is so despite the fact that the values of $l$ in
units of (either) string tension are in the range in which the
particle-like states were readily visible in $SO(4)$. We conclude that
if they are there, then the overlap must be suppressed by some power
of $k$ so that they have become invisible, within errors, for $SO(8)$.
That is to say, the particle-like state appears to decouple for both
large $l$ and large $k$.

\begin{figure}[htb]
\begin	{center}
\leavevmode
\input	{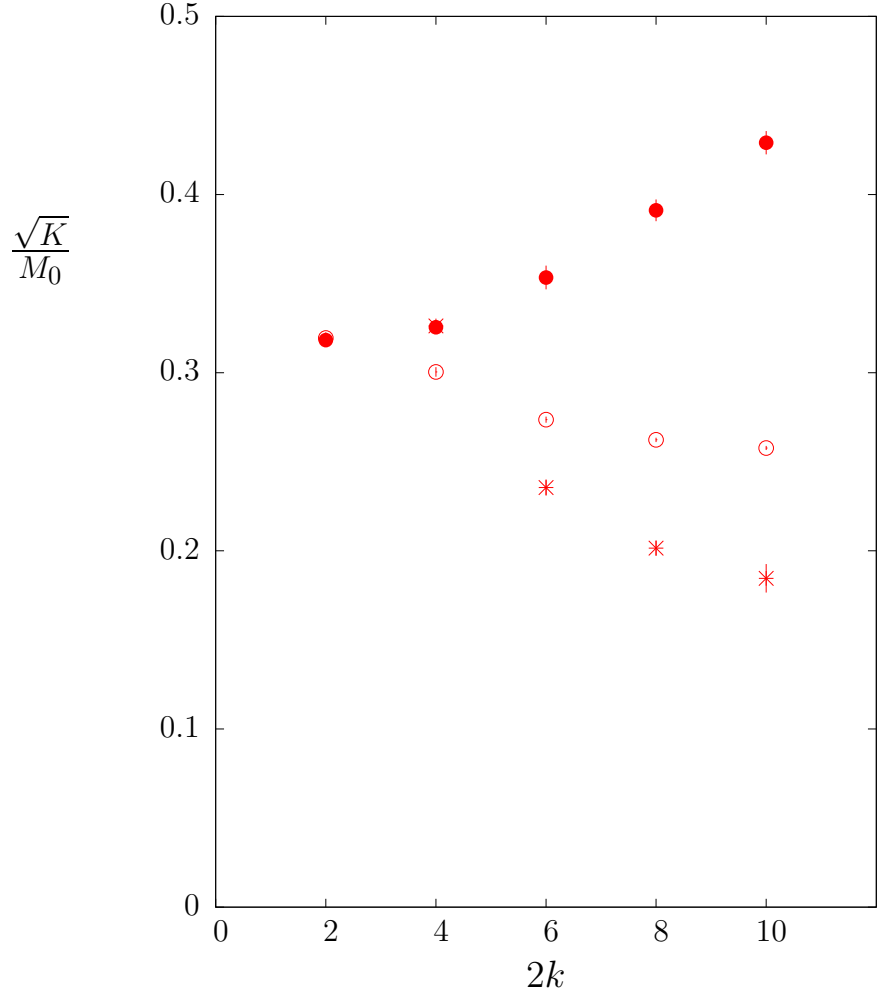}
\end	{center}
\caption{String tensions in $SO(2k)$ from the traces, $\circ$, and
  Pfaffians, $\bullet$, of flux loop operators in units of the mass gap.
  Also the Pfaffian string tension in units of the lightest Pfaffian
glueball, $\star$.}
\label{fig_KM0soN}
\end{figure}

The dependence of the string tension on $k$ is displayed
in Fig.\ref{fig_KM0soN}. We plot the two string tensions in units
of the lightest trace $J^{P}=0^{+}$ mass versus $2k$, just as we did
for the particle masses in Fig.\ref{fig_MPfM0}. The string tension
from the trace of flux loops decreases with $k$, which is no
surprise since one expects the  $SO(\infty)$ and  $SU(\infty)$
fundamental string tensions to be  equal while the  $SO(4)$ 
string tension is roughly twice that of $SU(2)$.
(All in units of the mass gap.) 
The string tension from the Pfaffian of the same flux loops
increases with $k$, in these units, with a nearly linear rise
for larger values of $k$. However in units of the lightest
Pfaffian particle the Pfaffian string tension also decreases
with increasing $k$, as we see in Fig.\ref{fig_MPfM0}.
From the coefficients of the linearly rising pieces (units
of the mass gap) we can estimate that this last ratio will
asymptote to a value $\sim 0.1$.

\section{Open questions}
\label{section_questions}

There are a number of issues that this study has brought out which
we have either only partially addressed or not addressed at all.
We briefly list some of them here. These are open problems which
need to be addressed.

We have seen that in $SO(4)$ the Pfaffian of a flux loop
operator can project onto a particle like state as well
as onto a state whose energy grows roughly linearly with
the length of the loop, but apparently with a string tension
that is larger than the $SU(2)$ one. To understand what these
states represent we need a more accurate determination of
their properties.

We have pointed out that one can generalise the $SO(2k)$ Pfaffian
to involve any $k$ adjoint operators, even if they differ. The
properties of such Pfaffians may be interesting. For example in $SO(4)$
an operator such as $\epsilon_{i_1i_2i_3i_4}\Phi_{l,i_1i_2}(x)\Phi_{ci_3i_4}(x)$
where $\Phi_l(x)$ is a non-contractible flux loop operator and
$\Phi_c(x)$ is a contractible loop should be exactly orthogonal
to all particle-like states, something that $\mathrm{Pf}\{\Phi_l\}$ itself
is not, as we have seen.

We have seen numerically that in $SO(6)$ if $\Phi_l$ is a non-contractible
loop operator then the operators  $\mathrm{Tr}\{\Phi_l\}$ 
and $\mathrm{Pf}\{\Phi_l\}$ correspond, within small errors, to the
$k=2A$ and $k=2S$ flux loops of $SU(4)$. This neat mapping cannot
however be exact since we know that in $SU(4)$ there is a small
but non-zero overlap between the $k=2A$ and $k=2S$ operators
\cite{2A2S_over},
whereas in $SO(6)$ the traced and Pfaffian operators are orthogonal.
Presumably it is the two $k=2$ orthogonal mixed states that correspond
to our two types of $SO(6)$ operators -- but this needs to be
better understood.

A more general question concerns the $SO(2N+1)$ gauge theories for
which we have no Pfaffian operator and about which we have had
nothing to say in this paper. On the other hand, earlier work has found
that the physics of the $SO(2N+1)$ and $SO(2N)$ gauge theories
seems to form one continuous family, even at small $N$
\cite{RLMT_SON_a,RLMT_SON_b}.
Although at large $N$ Pfaffian particles and strings become massive and
so decouple from the physics, at small $N$ they are relevant
and so one can ask whether there is something in the $SO(2N+1)$
theories that correspond to the Pfaffians in $SO(2N)$, e.g. in
$SO(3)$, which has the same Lie algebra as $SU(2)$.

\section{Conclusions}
\label{section_conclusions}


In this paper we showed that (generalised)  Pfaffian operators play an essential
role in completing the glueball spectrum calculations of $SO(2k)$
gauge theories. In particular for low $k$ where some of the  $SO(2k)$
gauge theories have the same Lie algebras as some $SU(N)$ gauge theories,
the Pfaffians provide the half of the spectrum that is missing when we
use only traces of loops for our operator basis. Thus they provide
the counterparts in $SO(6)$ of the $C=-$ particles in $SU(4)$, and
similarly in $SO(2)$ the counterparts of the $C=-$ particles in $U(1)$.
And they also provide the doubling of the spectrum in $SO(4)$ that one
might expect given that it has the same Lie algebra as $SU(2)\times SU(2)$.
As $k$ grows we could identify a linearly growing component to the mass,
which is no surprise given the fact that the Pfaffian of $SO(2k)$
contains pieces of the product of $k$ adjoint fields. The linear
growth of the lightest Pfaffian particle in units of the mass gap
was $M_{Pf}/M_{0^{++}} \simeq 0.06 + 0.47k$ which means that the
mass increases by $\sim 0.5 M_{0^{++}}$ as the Pfaffian operator length
increases by one adjoint field. It is intriguing that the same energy gap
(about one half of the mass gap) arises in other contexts, for example
in the massive excitation of the winding flux tube
\cite{SD_mass_a,SD_mass_b}.
In any case the fact that the masses of the Pfaffian particles increase
with $k$ means that they decouple as $k\to\infty$ and so do not upset
the expected equality of the $SO(\infty)$ and $SU(\infty)$ mass spectra.

We also investigated the states that couple to Pfaffians of the flux loop
operators whose trace projects onto confining flux tubes that wind around
a spatial circle (of our periodic lattice). We argued that these will
represent some kind of flux tube in $SO(4k+2)$ gauge theories but
not necessarily in $SO(4k)$ gauge theories. For $SO(2)$ we saw, rather
trivially, that the Pfaffian of flux loops corresponds to the $C=-$ flux
loop of $U(1)$, complementing the trace that corresponds to the 
$C=+$ flux loop of $U(1)$. Our calculations show
that for $SO(6)$ the energy increases nearly linearly with length as
one expects for a confining flux tube and it corresponds essentially
to the symmetric $k=2$ flux tube of $SU(4)$, complementing the trace
that corresponds to the antisymmetric $k=2$ flux tube of $SU(4)$.
This confirms our expectations for the  $SO(4k+2)$ series. 
$SO(4)$ is the first of the $SO(4k)$ series and here we found that
the  Pfaffian of flux loop operators had a visible but very weak overlap
onto a particle like state as well as a much larger overlap onto a
string state. The features of the latter are ambiguous within the
modest accuracy of our calculations. We also performed similar
calculations in $SO(8)$ which showed no sign of a particle-like state
which suggests some kind of large-$k$ decoupling. Together with
our $SO(10)$ calculations, all this showed that the $SO(2k)$ Pfaffian
string tension increases nearly linearly with $k$ at larger $k$,
so that just like the Pfaffian particles these Pfaffian strings will
decouple as $k\to\infty$.

Finally we emphasise that the present study has been very much an exploratory
one, although it has already succeeded in answering the question that
originally motivated it. At the trivial level this means that in various cases
the calculations
need greater accuracy and smaller errors in order to be completely convincing
and/or useful. Less trivially we have identified and discussed in  
Section~\ref{section_questions} a number of questions that need to
be addressed and a number of further calculations that need to be performed.

\section*{Acknowledgements}

The numerical computations were carried out on the computing cluster
in Oxford Theoretical Physics. The author is grateful to both Oxford
Theoretical Physics and to All Souls College for their support of this
research.

%
%
%
%


\newpage

\begin{table}[htb]
\begin{center}
\begin{tabular}{|c|c|cc|}\hline
  & $U(1)$, $\beta=2.2$ & \multicolumn{2}{|c|}{ $SO(2)$, $\beta=2.2$} \\ \hline
$J^{PC}$ &  Trace   &  Trace  &  Pfaffian \\ \hline
 $0^{--}$       & 0.2639(44) &          & 0.2630(26) \\
 $0^{--\star}$   & 0.726(22) &           & 0.740(13)  \\
 $0^{++}$       & 0.505(14) &  0.519(5) &   \\
 $0^{++\star}$   & 0.663(10) &  0.668(7) &   \\
 $0^{-+}$       & 1.11(5)   &  1.15(2)  &   \\
 $0^{+-}$       & 1.36(5)   &           & 1.36(1) \\
 $2^{++}$       & 0.711(6)  &  0.695(10)&     \\
 $2^{-+}$       & 0.831(10) &  0.821(7) &     \\
 $2^{--}$       & 0.937(13) &           & 0.930(12) \\
 $2^{+-}$       & 1.049(17) &           & 1.031(17) \\
 $1^{++}$       & 1.28(3)   & 1.29(3)   &     \\
 $1^{-+}$       & 1.24(4)   & 1.321(7)  &     \\
 $1^{--}$       & 1.093(16) &           & 1.099(13) \\
 $1^{+-}$       & 1.045(16) &           & 1.104(17) \\ \hline
 $l_f$         & 0.7492(36) & 0.7513(31) & 0.7452(49) \\ \hline
\end{tabular}
\caption{Lightest glueball masses in $U(1)$ and $SO(2)$ on a $28^236$ lattice at a coupling
  $\beta=2.2$ for various spins $J$ and parity $P$. For $U(1)$ also labelled by charge conjugation
  $C$, and for $SO(2)$ in the two sectors indicated as explained in text. Also shown is the
energy of the lightest fundamental flux tube, $l_f$, winding around a periodic spatial direction.}
\label{table_Mso2u1}
\end{center}
\end{table}

\begin{table}[htb]
\begin{center}
\begin{tabular}{|c|c|cc|}\hline
  & $SU(2)$, $\beta=13.87$ & \multicolumn{2}{|c|}{ $SO(4)$, $\beta=15.1$} \\ \hline
$J^{P}$ &  Trace   &  Trace  &  Pfaffian \\ \hline
 $0^{+}$       & 0.4807(11) &  0.4793(27) & 0.4782(21)  \\
 $0^{+\star}$   & 0.6851(50) &  0.710(5)   & 0.698(4)  \\
 $0^{-}$       & 1.002(7)   &  1.037(11)  & 0.972(28)  \\
 $2^{+}$       & 0.781(3)  &  0.800(6)    & 0.795(5)  \\
 $2^{-}$       & 0.781(5)  &  0.807(5)   & 0.791(6)     \\
 $1^{+}$       & 1.051(9)  & 1.138(53)   & 1.063(36)     \\
 $1^{-}$       & 1.066(10) & 1.170(15)  & 1.111(16)    \\
 $l_f$         & 0.5046(8)  & 1.028(7)  & 1.214(14) \\  \hline
\end{tabular}
\caption{Lightest glueball masses $SU(2)$ and $SO(4)$ on a $50^256$ lattice at the couplings
  shown for various spins $J$ and parity $P$.   Also shown is the
energy of the lightest fundamental flux tube, $l_f$, winding around a periodic spatial direction.}
\label{table_Mso4su2}
\end{center}
\end{table}

\begin{table}[htb]
\begin{center}
\begin{tabular}{|c|c|cc|}\hline
  & $SU(2)$, $\beta=13.87$ & \multicolumn{2}{|c|}{ $SO(4)$, $\beta=15.1$} \\ \hline
$J^{P}$ &  Trace   &  Trace  &  Pfaffian \\ \hline
 $0^{+}$       & 0.4777(12)  &  0.4751(30) & 0.4787(20)  \\
 $0^{-}$       & 1.002(10)   &  1.047(11)  & 1.037(10)  \\
 $2^{+}$       & 0.7441(10)  &  0.764(9)   & 0.735(9)  \\
 $2^{-}$       & 0.7909(22)  &  0.798(5)   & 0.784(12)     \\
 $1^{+}$       & 1.069(6)    & 1.084(39)   & 1.075(33)     \\
 $1^{-}$       & 1.075(3)    & 1.176(16)   & 1.056(46)    \\ \hline
 $l_f$         & 0.5046(8)    & 0.6791(29)  & 0.549(22) \\  \hline
\end{tabular}
\caption{Lightest glueball masses $SU(2)$ and $SO(4)$ on a $34^256$ lattice at the couplings
  shown for various spins $J$ and parity $P$.   Also shown is the
energy of the lightest flux tube, $l_f$, winding around a periodic spatial direction.}
\label{table_Mso4su2_small}
\end{center}
\end{table}

\begin{table}[htb]
\begin{center}
\begin{tabular}{|c|c|cc|}\hline
  & $SU(4)$, $\beta=59.14$ & \multicolumn{2}{|c|}{ $SO(6)$, $\beta=46.0$} \\ \hline
$J^{PC}$ &  Trace   &  Trace  &  Pfaffian \\ \hline
 $0^{++}$       & 0.4605(33) &  0.4612(19) &   \\
 $0^{++\star}$   & 0.689(9)   &  0.708(5)  &   \\
 $0^{--}$       & 0.683(3)   &            & 0.692(4) \\
 $0^{--\star}$   & 0.857(5)   &            & 0.874(7)  \\
 $0^{-+}$       & 0.974(10)  &  1.013(11) &   \\
 $0^{+-}$       & 1.084(14)  &            & 1.179(21) \\
 $2^{++}$       & 0.770(4)   &  0.759(15) &     \\
 $2^{-+}$       & 0.771(4)   &  0.764(15) &     \\
 $2^{--}$       & 0.919(6)   &            & 0.913(21) \\
 $2^{+-}$       & 0.917(8)   &            & 0.935(8) \\
 $1^{++}$       & 1.031(39)  & 1.126(14)  &     \\
 $1^{-+}$       & 1.083(5)   & 1.087(44)  &     \\
 $1^{--}$       & 1.048(9)   &            & 1.094(15) \\
 $1^{+-}$       & 1.062(5)   &            & 1.026(54) \\ \hline
 $l_{k=1}$      & 0.5389(11) & 0.7138(89) & 1.166(55) \\  \hline
\end{tabular}
\caption{Lightest glueball masses in $SU(4)$ and $SO(6)$ on a $46^248$ lattice at the couplings 
  shown for various spins $J$ and parity $P$. For $SU(4)$ labelled by charge conjugation
  $C$, and for $SO(6)$ in the two sectors indicated, as explained in text. Also shown is the
energy of the lightest $k$-string, $l_k$, winding around a periodic spatial direction.}
\label{table_Mso6su4}
\end{center}
\end{table}

\begin{table}[htb]
\begin{center}
\begin{tabular}{|c|c|cc|}\hline
  & $SU(4)$, $\beta=59.14$ & \multicolumn{2}{|c|}{ $SO(6)$, $\beta=46.0$} \\ \hline
$J^{PC}$ &  Trace   &  Trace  &  Pfaffian \\ \hline
 $0^{++}$       & 0.4617(26) &  0.4639(16) &   \\
 $0^{++\star}$   & 0.698(3)   &  0.706(5)  &   \\
 $0^{--}$       & 0.680(4)   &            & 0.677(4) \\
 $0^{--\star}$   & 0.849(8)   &            & 0.872(9)  \\
 $0^{-+}$       & 0.967(18)  &  0.957(30) &   \\
 $0^{+-}$       & 1.114(11)  &            & 1.142(21) \\
 $2^{++}$       & 0.772(3)   &  0.778(5)  &     \\
 $2^{-+}$       & 0.777(3)   &  0.773(6)  &     \\
 $2^{--}$       & 0.923(3)   &            & 0.945(10) \\
 $2^{+-}$       & 0.923(3)   &            & 0.947(9) \\
 $1^{++}$       & 1.087(17)  & 1.096(14)  &     \\
 $1^{-+}$       & 1.085(5)   & 1.129(16)  &     \\
 $1^{--}$       & 0.988(38)  &            & 1.083(16) \\
 $1^{+-}$       & 1.048(15)  &            & 1.080(15) \\ \hline
 $l_{k=1}$      & 0.4149(10) &            &  \\ 
 $l_{k=2A}$     & 0.5681(14) & 0.5642(19) &    \\ 
 $l_{k=2S}$     & 0.986(14)  &            & 0.977(26) \\ \hline
\end{tabular}
\caption{Lightest glueball masses in $SU(4)$ and $SO(6)$ on a $36^244$ lattice at the couplings 
  shown for various spins $J$ and parity $P$. For $SU(4)$ labelled by charge conjugation
  $C$, and for $SO(6)$ in the two sectors indicated, as explained in text. Also shown is the
energy of the lightest $k$-string, $l_k$, winding around a periodic spatial direction.}
\label{table_Mso6su4_small}
\end{center}
\end{table}

\begin{table}[htb]
\begin{center}
\begin{tabular}{|c|cc|cc|}\hline
  & \multicolumn{2}{|c|}{ $SO(8)$, $\beta=84.0$} & \multicolumn{2}{|c|}{ $SO(10)$, $\beta=120.0$} \\ \hline
$J^{P}$ &  Trace  &  Pfaffian  &  Trace  &  Pfaffian \\ \hline
 $0^{+}$       & 0.5101(21) & 0.990(15)   &  0.6022(24)  &  1.400(57)  \\
 $0^{+\star}$   & 0.768(6)   & 1.182(25)   &  0.9210(42)  &  1.73(14)  \\
 $0^{-}$       & 1.069(19)  & 1.528(73)   &  1.263(35)   &  2.206(55)  \\
 $2^{+}$       & 0.848(7)   & 1.276(30)   &  1.001(6)    &  1.826(30)  \\
 $2^{-}$       & 0.856(5)   & 1.270(24)   &  1.009(5)    &  1.841(27)  \\ 
 $1^{+}$       & 1.194(21)  & 1.429(36)   &  1.455(15)   &  2.11(6)    \\
 $1^{-}$       & 1.188(24)  & 1.356(46)   &  1.476(13)   &  2.09(6)    \\  \hline
 $l_f$         & 0.4829(15) & 1.126(11)   &  0.5057(17)  &  1.445(44)  \\  \hline
 \end{tabular}
\caption{Lightest glueball masses in $SO(8)$ on a  $28^236$ lattice  and in $SO(10)$ on
  a $22^230$ lattice at the couplings shown for various spins $J$ and parity $P$. 
  Also shown is the energy of the lightest flux loop, $l_f$, winding around a periodic spatial direction.}
\label{table_Mso8so10}
\end{center}
\end{table}


\begin{table}[htb]
\begin{center}
\begin{tabular}{|cc|c|cc|cc|}\hline
  group & $\beta$ & $l$ & $aE_{Tr}(l)$  & $a\surd\sigma_{Tr}$ & $aE_{Pf}(l)$  & $a\surd\sigma_{Pf}$ \\ \hline
SO(2)  &  2.2  &  28  & 0.7513(31) & 0.1658(4)  & 0.7452(49) & 0.1652(5)    \\  \hline
SO(4)  & 15.1  &  50  & 1.0284(68) & 0.1442(5)  & 1.214(14)  & 0.1565(9)    \\ 
       &       &  46  & 0.913(18)  & 0.1418(14) & 1.121(15)  & 0.1569(11)   \\
       &       &  42  & 0.8706(33) & 0.1450(3)  & 0.971(12)  & 0.1530(10)   \\  \hline
SO(6)  & 46.0  &  46  & 0.7138(89) & 0.1256(8)  & 1.166(55)  & 0.1600(37)   \\ 
       &       &  36  & 0.5642(19) & 0.1268(3)  & 0.977(26)  & 0.1660(23)   \\  \hline
SO(8)  & 84.0  &  28  & 0.4829(15) & 0.1339(2)  & 1.126(11)  & 0.2022(10)   \\ 
       &       &  22  & 0.3684(15) & 0.1336(3)  & 0.824(17)  & 0.1963(20)   \\  \hline
SO(10) & 120.0 &  22  & 0.5057(17) &  0.1552(3) & 1.445(44)  & 0.2584(38)   \\  \hline
 \end{tabular}
\caption{Energy and resulting string tension of lightest flux loop of length $l$ in the Trace and Pfaffian
  sectors for the groups shown and for our largest valuesof $l$. The string tensions have been extracted
  using the simple `Nambu-Goto' formula in eqn(\ref{eqn_NG}).}
\label{table_KsoN}
\end{center}
\end{table}

\begin{table}[htb]
\begin{center}
\begin{tabular}{|c|c|cc|cc|}\hline
  &  \multicolumn{3}{|c|}{$SU(4)$, $\beta=59.14$} & \multicolumn{2}{|c|}{ $SO(6)$, $\beta=46.0$} \\ \hline
$l.l_\perp.l_t$ & $aE_{k=1}$ & $aE_{k=2A}$ & $aE_{k=2S}$ & $aE_{k=1}^{Tr}$ & $aE_{k=1}^{Pf}$   \\  \hline
 18.48.64 & 0.1812(9)  & 0.2572(20)  & 0.427(8)  & 0.2546(16)  & 0.426(12)  \\  
 22.42.52 & 0.2379(5)  & 0.3316(15)  & 0.561(8)  & 0.3286(12)  & 0.569(8)   \\  
 26.40.52 & 0.2911(7)  & 0.3948(35)  & 0.690(7)  & 0.3964(19)  & 0.695(9)   \\  
 30.40.50 & 0.3416(10) & 0.4694(10)  & 0.795(19) & 0.4660(19)  & 0.803(17)  \\  
 36.36.44 & 0.4142(17) & 0.5670(23)  & 0.986(14) & 0.5642(19)  & 0.977(26)  \\  
 46.46.48 & 0.5389(11) & 0.7331(33)  & 1.23(5)   & 0.7138(89)  & 1.17(6)    \\   \hline
 \end{tabular}
\caption{Energy of lightest flux loop of length $l$ in the Trace and Pfaffian
  sectors for $SO(6)$ and in the $k=1$, $k=2A$ and $k=2S$ sectors for $SU(4)$.}
\label{table_Elso6su4}
\end{center}
\end{table}

\begin{table}[htb]
\begin{center}
\begin{tabular}{|c|cc|ccc|}\hline
  &  \multicolumn{2}{|c|}{$SU(2)$, $\beta=59.14$} & \multicolumn{3}{|c|}{ $SO(4)$, $\beta=15.1$} \\ \hline
$l.l_\perp.l_t$ & $aE_{k=1}$ & $aE_{eff}$ & $aE^{Tr}$ & $aE^{Pf}$ & $a\tilde{E}^{Pf}$   \\  \hline
 18.38.56 & 0.1550(7)  & 0.3433(14) & 0.3094(39)  &   0.3529(27) & [0.527(9)]   \\
 30.30.56 &            &            & 0.5874(42)  &   0.589(17) &  0.572(27)   \\
 34.34.56 & 0.3351(6)  & 0.6861(13) & 0.6791(29)  &   0.679(11) &  0.549(22)   \\
 38.38.56 &            &            & 0.7694(54)  &   0.755(37) &  0.582(27)   \\
 42.42.56 &            &            & 0.8706(33)  &   0.971(12) &  [0.48(14)]   \\
 46.46.48 &            &            & 0.913(18)   &   1.121(15) &  [0.88(10)]   \\
 50.50.56 & 0.5046(8)  & 1.0198(16) & 1.0284(68)  &   1.214(14) &  [0.46(18)]   \\   \hline
 \end{tabular}
\caption{Energy of lightest flux loop of length $l$ in the Trace and Pfaffian
  sectors for $SO(4)$ and in the fundamental for $SU(2)$. $E_{eff}$ is the energy
  the $SU(2)$ flux loop would have if the string tension was twice as large.
  See text for explanation of Pfaffian energies.}
\label{table_Elso4su2}
\end{center}
\end{table}

\begin{figure}[htb]
\begin	{center}
\leavevmode
\input	{plot_EleffPfso6.tex}
\end	{center}
\caption{Effective energies of the ground state Pfaffian string
  in $SO(6)$ at $\beta=46.0$ for lengths $l=18,22,26,30,36,46$ in
  order. Pairs of lines are $\pm 1\sigma$ around mean of our
  asymptotic energy estimates.}
\label{fig_EleffPfso6}
\end{figure}

\begin{figure}[htb]
\begin	{center}
\leavevmode
\input	{plot_EleffTrso6.tex}
\end	{center}
\caption{Effective energies of the ground state traced string
  in $SO(6)$ at $\beta=46.0$ for lengths $l=18,22,26,30,36,46$ in
  order. Pairs of lines are $\pm 1\sigma$ around mean of our
  asymptotic energy estimates.}
\label{fig_EleffTrso6}
\end{figure}

\begin{figure}[htb]
\begin	{center}
\leavevmode
\input	{plot_EleffTrso4.tex}
\end	{center}
\caption{Effective energies of the ground state traced string
  in $SO(4)$ at $\beta=15.1$ for lengths $l=18,30,34,38,42,46,50$ in
  order. Pairs of lines are $\pm 1\sigma$ around mean of our
  asymptotic energy estimates.}
\label{fig_EleffTrso4}
\end{figure}

\begin{figure}[htb]
\begin	{center}
\leavevmode
\input	{plot_EleffPfso4.tex}
\end	{center}
\caption{Effective energies of the ground state Pfaffian string
  in $SO(4)$ at $\beta=15.1$ for lengths $l=18,30,34,38,42,46,50$ in
  order. Pairs of lines are $\pm 1\sigma$ around mean of our
  asymptotic energy estimates.}
\label{fig_EleffPfso4}
\end{figure}

\begin{thebibliography}{99}

\bibitem{RLMT_SON_a}
R. Lau amd M. Teper, {\it SO(N) gauge theories in 2+1 dimensions: glueball spectra and confinement},
JHEP 1710 (2017) 022 [arXiv:1702.03717].

\bibitem{RLMT_SON_b}
R. Lau amd M. Teper, {\it The deconfining phase transition of SO(N) gauge theories in 2+1 dimensions},
JHEP 1603 (2016) 072 [arXiv:1510.07841].

\bibitem{MT_SON}
M. Teper {\it SO(4), SO(3) and SU(2) gauge theories in 2+1 dimensions: comparing glueball spectra and string tensions},
arXiv:1801.05693.

\bibitem{AAMT_SUN}
A. Athenodorou and M. Teper, {\it SU(N) gauge theories in 2+1 dimensions: glueball spectra and k-string tensions},
JHEP 1702 (2017) 015 [arXiv:1609.03873].  

\bibitem{largeN_SUN}
  G. 't Hooft, {\it A Planar Diagram Theory for Strong Interactions}, Nucl. Phys. B72 (1974) 461.

\bibitem{largeN_SON}
  C. Lovelace, {\it Universality at large N}, Nucl. Phys. B201 (1982) 333.

\bibitem{MT_spinorial}
  M. Teper, {\it Spinorial flux tubes in SO(N) gauge theories in 2+1 dimensions},
  arXiv:1712.01185.
  
\bibitem{EW_98}
  E. Witten, {\it Baryons and branes in anti-de Sitter space },
  JHEP 9807 (1998) 006 [arXiv:hep-th/9805112]. 

\bibitem{block_a}
  M. Teper, {\it An improved method for lattice glueball calculations},
  Phys.Lett. B183 (1987) 345.

\bibitem{block_b}
  M. Teper, {\it SU(N) gauge theories in 2+1 dimensions},
  Phys.Rev. D59 (1999) 014512 [arXiv:hep-lat/9804008].

\bibitem{HMMT_a}
  H. Meyer and M. Teper, {\it High spin glueballs from the lattice},
  Nucl.Phys. B658 (2003) 113 [arXiv:hep-lat/0212026].
  
\bibitem{HMMT_b}
  H. Meyer, {\it Glueball Regge trajectories},
  D.Phil Thesis, University of Oxford, 2005 [arXiv:hep-lat/0508002].

\bibitem{string_LAT_a}
  A. Athenodorou, B. Bringoltz and M. Teper, {\it Closed flux tubes and their string description in D=2+1 SU(N) gauge theories},
  JHEP 1105 (2011) 042 [arXiv:1103.5854].
  
\bibitem{string_LAT_b}
  A. Athenodorou and M. Teper, {\it Closed flux tubes in D=2+1 SU(N) gauge theories: dynamics and effective string description},
  JHEP 1610 (2016) 093 [arXiv:1602.07634].
    
\bibitem{string_TH_a}
  O. Aharony and Z. Komargodski, {\it The Effective Theory of Long Strings},
  JHEP 1305 (2013) 118 [arXiv:1302.6257].
  
\bibitem{string_TH_b}
  O. Aharony and E. Karzbrun, {\it On the effective action of confining strings},
  JHEP 0906 (2009) 012 [arXiv:0903.1927].
  
\bibitem{string_TH_c}
  S. Dubovsky, R. Flauger and V. Gorbenko, {\it Effective String Theory Revisited},
  JHEP 1209 (2012) 044 [arXiv:1203.1054].

\bibitem{U1_conf_a}
  A. Polyakov, {\it Quark Confinement and Topology of Gauge Groups},
  Nucl.Phys. B120 (1977) 429.
  
\bibitem{U1_conf_b}
  A. Polyakov, {\it Gauge fields and strings} (Harwood Academic Publishers, 1987).
  
\bibitem{U1_conf_c}
  R. Wensley and J. Stack, {\it Monopoles and Confinement in Three-dimensions},
  Phys.Rev.Lett. 63 (1989) 1764.
  
\bibitem{U1_conf_d}
  R. Wensley, {\it Monopoles and U(1) lattice gauge theory},
  Ph.D. Thesis, University of Illinois, 1989.
  
\bibitem{U1_conf_e}
  T. Copeland {\it Monopoles and confinement in U(1) lattice gauge theory},
  D.Phil Thesis, University of Oxford, 1990.

\bibitem{BLMT}
  B. Lucini and M. Teper, {\it Confining strings in SU(N) gauge theories},
  Phys.Rev. D64 (2001) 105019 [arXiv:hep-lat/0107007].
  
\bibitem{2A2S_over}
  A. Athenodorou, B. Bringoltz and M. Teper, {\it On the spectrum of closed k=2 flux tubes in D=2+1 SU(N) gauge theories},
  JHEP 0905 (2009) 019 [arXiv:0812.0334].
  
\bibitem{SD_mass_a}
  S. Dubovsky, R. Flauger and V. Gorbenko, {\it Evidence for a new particle on the worldsheet of the QCD flux tube},
  Phys.Rev.Lett. 111 (2013) 062006 [arXiv:1301.2325].
  
\bibitem{SD_mass_b}
  A. Athenodorou and M. Teper, {\it Closed flux tubes in higher representations and their string description in D=2+1 SU(N) gauge theories},
  JHEP 1306 (2013) 053 [arXiv:1303.5946].




\end{thebibliography}
\end{document}